\documentclass[prc,amsmath,twocolumn,superscriptaddress,showpacs,nofootinbib]
{revtex4} 

\usepackage{graphicx}
\usepackage[hyperfootnotes=false]{hyperref}
\usepackage{color} 

\def\sc{{\rm sc}}
\def\tsc{\tau_{\rm sc}\,}
\def\tac{\tau_{\rm ac}\,}

\def\tcross{\tau_{\rm cross}\,}
\def\tesc{T_{\rm esc}}
\def\tloss{\tau_{\rm L}\,}

\begin{document}

\title{Determination of Acceleration Mechanism Characteristics Directly and
Non-Parametrically from Observations: Application to Supernova Remnants}
\author{Vah\'e Petrosian}
\email{vahep@stanford.edu}
\affiliation{Department of Physics and KIPAC, Stanford University, Stanford, CA
94305, USA}
\affiliation{Department of Applied Physics, Stanford University, Stanford, CA
94305, USA}
\author{Qingrong Chen}
\affiliation{Department of Physics and KIPAC, Stanford University, Stanford, CA
94305, USA}


\begin{abstract}

We have developed an inversion method for determination of the characteristics
of the acceleration
mechanism directly and non-parametrically from observations, in contrast to
the usual forward
fitting of parametric model variables to observations. In two recent papers 
\cite{Petrosian10, ChenQ13},
we demonstrate the efficacy of this inversion method by its application to
acceleration of electrons
in solar flares based on stochastic acceleration by turbulence. Here we explore
its application for
determining the characteristics of shock acceleration in supernova remnants
(SNRs) based on the electron spectra deduced from the 
observed nonthermal radiation from SNRs and the spectrum of the cosmic ray 
electrons
observed near the Earth. These spectra  are related by the
process of escape of the
electrons from SNRs and  energy loss during  their transport  in the galaxy.
Thus,
these observations allow us to determine spectral characteristics of the
momentum and pitch angle
diffusion coefficients, which play crucial roles in both direct acceleration by
turbulence and in
high Mach number shocks. Assuming that the average electron
spectrum
deduced from a few well known SNRs is representative of those in the solar
neighborhood we find interesting discrepancies between our deduced forms for
these
coefficients and those expected from well known wave-particle interactions. This
may indicate that
the standard assumptions made in treatment of shock acceleration  need
revision. In particular, the
escape of particles from SNRs may be more complex than generally assumed.

\end{abstract}

\maketitle

\section{Introduction}
\label{intro}

Acceleration of charge particles  in the universe happens on
scales from planetary magnetospheres to clusters of galaxies and at energies
ranging  from nonrelativistic values to $>$10$^{19}$ eV ultra high energy cosmic
rays (UHECRs). The
particles are
observed directly as cosmic rays (CRs), solar energetic particles, 
or indirectly by their interactions with background
matter and electromagnetic fields (magnetic fields and photons), which
give rise to heating and  ionization of the plasma, and  nonthermal radiation
extending from long wavelength radio to $>$TeV gamma-rays. In spite of more than
a
century of observations, the exact mechanism of acceleration is still being
debated and the detailed model parameters are poorly constrained. Clearly
electric
fields are involved in any acceleration mechanism. 
 Large scale
electric fields  have been found to be important  in some unusual astrophysical
sources such as 
magnetospheres of neutron stars (pulsars and perhaps magnetars) and  in
so-called double-layers. However, here we are interested in commonly
considered mechanisms  based on
the original Fermi process \cite{Fermi49}, which involves scattering of particles by
fluctuating
electric and magnetic fields (or plasma turbulence) or converging flows as in
shocks.  

The usual approach of determining the acceleration model and its characteristics
is to use
the forward fitting (FF) method, whereby the model particle spectra based on an
assumed
mechanism and some parametric form of its characteristics are fitted to
observations. For radiating sources,  FF is carried out in two stages, first
fitting the photon spectra to an assumed radiation mechanism from a
parametrized particle spectrum, then fitting the latter to the acceleration
model. This approach, even though one can
never be certain of the uniqueness of the results, has been fairly successful,
and for some observations, e.g., those with poorly  determined spatially
unresolved spectra, 
is the best one can do. But in sources with  richer observations one can do
better. 

In this paper we present a new approach which allows a non-parametric
determination of acceleration parameters, mainly their energy dependence,
irrespective of some of the details of the acceleration mechanism, directly from
the observed radiation or otherwise deduced particle spectra. This is done by
the {\it inversion} 
of the kinetic differential equations describing the particle acceleration and
transport. In our
first
paper on this subject
\cite{Petrosian10},
we applied this technique to inversion of hard X-ray images of solar
flares from the Reuven Ramaty High Energy Solar Spectroscopic Imager
({\it RHESSI}) and determined the energy dependence of the escape time from the
acceleration region and from it the energy dependence of the rate of scattering
of the
particles, presumably due to plasma turbulence, which
is related to the pitch angle diffusion
coefficient $D_{\mu\mu}$, where $\mu$ is the cosine of the pitch angle. In a
more recent paper \cite{ChenQ13}, we have shown that from the same
data we can also determine the  energy diffusion coefficient $D_{\rm EE}$, which
is
related to the momentum diffusion coefficient $D_{pp}$. In both papers
we formulated this in the framework of stochastic acceleration (SA) by
plasma waves or turbulence, which is same as the original Fermi process,
nowadays referred to as second-order Fermi
acceleration process. Here we extend this approach to
simultaneous determination of  the scattering and acceleration rates, which
depend primarily on  $D_{\mu\mu}$ and
$D_{pp}$, to situations where both  SA
by turbulence and  acceleration by a shock play important roles. As in previous
papers
we carry this out in the framework of the so called
leaky box model. In the next section we present the
kinetic equation
describing both acceleration processes,  and in \S \ref{inv} we
describe the process of the inversion and the required data for it. In \S
\ref{sec_SNR} we describe
possible application of this method
to
the acceleration of electrons in supernova remnants (SNRs). Interpretation
and
discussions of
the results are shown in \S
\ref{sec_interp} and a brief summary is presented in \S \ref{sec_sum}.

\section{Kinetic Equations and the Leaky Box Model}
\label{model}

The discussion below is a brief summary of this subject given in a recent review
by \cite{Petrosian12} describing the conditions under which the so-called
leaky-box model is a good approximation. As emphasized in this review, and
recognized by
the community at large, it is clear now that plasma waves or turbulence play an
essential role in the acceleration of charged particles 
in a variety of magnetized astrophysical and space environments.
Turbulence is expected to be produced by large scale  flows in most
astrophysical situations because of the prevailing large Reynolds numbers. Once
generated
on a scale $L$ comparable to the size of the source it undergoes dissipationless
cascade from
large to small
spatial scales, or from small  wave numbers  $k_{\rm min} \sim 2\pi/L$
up to the dissipation scale given by $k_{\rm max}$, generally with a power law
energy density
distribution $W(k)\propto k^{-q}$. 
Resonant interactions between particles and
small amplitude electromagnetic fluctuations of turbulence
cause diffusion of particles in the phase space. For magnetized plasmas this
process can be described by the Fokker-Planck (FP) kinetic equation for
gyro-phase averaged, four dimensional (4-D)  particle distribution function
$f(t, \mu, p,
s)$, where $s$ is the distance along the magnetic
field lines. This equation involves, in addition to  $D_{\mu\mu}$ and $D_{pp}$,
a third coefficient $D_{\mu p}=D_{p \mu}$,%
\footnote{All three coefficients depend on $p$ and $\mu$ and are $\propto
\Omega
f_{\rm turb}$, where $\Omega$ is
the particle gyro frequency and $f_{\rm turb}=(\delta B/B)^2$ is the ratio of
the turbulent to total magnetic field energy densities (see e.g.
\cite{Pryadko97}.}
as well as a source term ${\dot
{\cal
S}}(t,\mu,p,s)$ 
and energy losses or gains due
to interactions of particles with background plasma (with density $n$,
temperature
$T$, magnetic field $B$ and soft photon energy density $u_{\rm ph}$). These
interactions cause 
stochastic acceleration, e.g., \cite{Sturrock66b, Schlickeiser89}, 
in which particles systematically gain energy with a rate that is proportional
to
the square of the wave-to-particle velocity ratio as in the  second-order Fermi
process. 

Also shown in \cite{Schlickeiser89}, the 4-D differential equation can be
reduced to a
3-D equation, when the scattering time
$\tsc \sim 1/D_{\mu\mu}$ is shorter than the dynamic time $\tau_{\rm dyn}$ and
the
crossing time $\tcross \sim L/v$.%
\footnote{Note that here $v$ is the particle velocity and in what follows the
size $L$  refers to the length
of the bundle of magnetic lines the particles are tied to. For chaotic fields
this could be much larger than the physical size of the turbulent acceleration
region.}
Then the momentum distribution is nearly isotropic and
one can define the pitch angle averaged quantities,  $F(t, p,
s)=\frac{1}{2}\int_{-1}^{+1} f(t,
\mu, p, s)d\mu$ and ${\dot S}(t, p, s)=\frac{1}{2}\int_{-1}^{+1} {\dot {\cal
S}}(t, \mu, p, s)d\mu$,
{and use three pitch angle-averaged
transport coefficients 
\begin{eqnarray}
\kappa_{ss}&=& (v/2)^2\langle(1-\mu^2)^2/D_{\mu\mu}\rangle\,,
\label{kss}\\
\kappa_{sp}&=& v/(2p)\langle(1-\mu^2)D_{\mu p}/D_{\mu\mu}\rangle\,,
\label{ksp}\\
\kappa_{pp} &=& \langle D{pp}-D^2_{\mu p}/D_{\mu\mu}\rangle/p^2\,.
\label{kpp}
\end{eqnarray}
(see \cite{Petrosian12}) to describe spatial and momentum diffusion rates.}
\cite{Schlickeiser89} and others,
in most subsequent applications of this equation, were interested in
acceleration
by Alfv\'en waves (with velocity $v_A$), in which case the diffusion
coefficients are related as $D_{\mu\mu}:D_{\mu
p}/p:D_{pp}/p^2=1:(v_A/v):(v_A/v)^2$. Limiting their analysis to  low
magnetization and high energy
particles, i.e. for  $v_A/v\ll 1$ they used the inequities  $D_{\mu\mu}\gg
D_{\mu p}/p\gg D_{pp}/p^2$ to obtain the simplified equation.
However, as was pointed out by \cite{Pryadko97}, at low energies
and for
strong magnetic fields,  other plasma waves become more important than the
Alfv\'en waves and 
these inequalities are no longer valid,  e.g., \cite{Dung94}. \cite{Pryadko97} 
suggested another approximation for the FP equation for the opposite limit, 
$D_{pp}/p^2\gg D_{\mu p/p}\gg D_{\mu\mu}$, in which case the momentum diffusion 
is the dominant term. These ideas were further developed by \cite{Petrosian04} 
and
summarized in \cite{Petrosian12}. It turns out that if again $\tsc \ll
\tcross$  and $\tau_{\rm dyn}$, then this situation can be described by  the
same 3-D equation with slightly different coefficients. (The proof of this
assertion will be
presented elsewhere.) 

Finally a second simplification can be used for both  cases if the
acceleration region is homogeneous, or if one deals with a spatially unresolved
acceleration region where one is interested in spatially integrated equations.
In this case
it is convenient to 
define  the 2-D distribution function in terms of the particle energy $E$,
$N(t,E)dE=\int
dV[4\pi p^2F(t,s,p)dp]$  and  ${\dot Q}_{\rm inj}(t,E)dE=\int
dV[4\pi p^2{\dot S}(t,s,p)dp]$, introduce spatially averaged terms 
${\bar X}=\int X(s)F(s)ds/\int F(s)ds$ and replace the  spatial 
diffusion term  by an escape
term. Then  we obtain the following well known equation, sometimes referred to
as the {\it
leaky box model},
\begin{align}
\label{lbox}
{\partial N \over \partial t}
 = & {\partial \over \partial E} \left[D_{\rm EE}{\partial N\over \partial E}
\right]
 - {\partial \over \partial E} \left[(A(E) - \dot E_{\rm L}) N\right]
\nonumber\\
 & - {N \over T_{\rm esc}} +{\dot Q}_{\rm inj}\,,
\end{align}
where $D_{\rm EE}=v^2p^2\bar {\kappa}_{pp}$,  $A(E)$ and $\dot E_{\rm L}$ are
the
direct
acceleration and energy loss rates, and ${\dot Q}_{\rm inj}(t, E)$ and
$N(t,E)/T_{\rm
esc}$
represent the rates
of injection and escape of particles in and out of the whole acceleration site.%
\footnote{This clearly is an approximation with the primary assumption being
that the transport coefficients have a slow  spatial variation. See
\cite{Petrosian12} for
details.}
For purely  SA, the direct acceleration rate%
\footnote{In another, more standard form of the kinetic equation
\cite{Chandrasekhar43}, the first three terms for stochastic acceleration
(without ${\dot E}_L$) are written as
$\frac{\partial N}{\partial t}=\frac{\partial^2(D_{\rm EE}N)}{\partial E^2}
-\frac{\partial(\tilde{A}N)}{\partial E}$, 
where ${\tilde A}(E)=\frac{D_{\rm EE}}{E}\xi + \frac{d D_{\rm EE}}{dE}$ gives
the direct energy gain rate.
Defining the total energy of the accelerated particles as 
${\cal E}(t)=\int_0^\infty EN(E, t)dE$, it is straightforward to show that
integration of the above
equation over energy gives 
$\frac{d{\cal E}}{dt} = \int_0^\infty \tilde{A}(E)N dE$
\cite{Tsytovich77}, showing that $\tilde{A}(E)$ provides a more accurate
representation of  the direct energy gain rate than $A(E)$. In what follows we
use the form given
in Eq. (\ref{lbox}) which is more convenient for the inversion procedure.}
\begin{align}
\label{ASA}
A_{\rm SA}(E)= 2{\bar \xi}D_{\rm EE}/E=2\xi' E\bar {\kappa}_{pp},
\end{align}
where
\begin{align}
{\bar \xi}={\gamma^2-0.5\over \gamma^2+\gamma}\,\,\,\, {\rm and}\,\,\,\,\,
\xi' = \frac{(\gamma+1)(2\gamma^2-1)}{2\gamma^3}.
\end{align}
The term $\xi'$  is  nearly equal to 1 at all $\gamma$ (it has a maximum of
$\sim$1.3 for $\gamma\sim 1.8$).%
\footnote{Here ${\bar \xi}=\xi/2$, where $\xi$ is used in \cite{Petrosian12} and our
earlier papers.}

Because the acceleration rate in stochastic acceleration is
proportional to the square of the velocity ratio $v_{A}/v$, it is often
regarded to be too slow
to account for production of high-energy particles, 
especially in comparison to acceleration in a shock
(or a converging flow in general). For a shock   with velocity $u_{\rm sh}$, 
a particle of velocity $v$ upon crossing it gains  momentum linearly with
velocity; $\delta p= p(u_{\rm sh}/v)$, and therefore this often is referred to
as
a first-order Fermi process. There are several
misconceptions associated with the above statement. The first is that the
diffusion
coefficients, in general, increase with decreasing particle energy so that SA
can be very efficient in the acceleration of low energy particles
in the background plasma,
which is
where all acceleration processes must start \cite{Hamilton92, Dung94}. The
second is that shock acceleration is not related to the original 
\underline{first-order Fermi} process \cite{Fermi54},   
and the third is  that shock acceleration rate is also second
order.

In an unmagnetized shock, or in a  shock
with  magnetic field parallel to the  shock velocity,
acceleration requires
an scattering agent to recycle particles repeatedly across the shock. Turbulence
is the most likely agent for this. The acceleration rate then is $\delta
p/\delta t$, where  the recycling time $\delta t\sim
{\bar \kappa}_{ss}/(vu_{\rm
sh})$ 
\cite{Krymsky79, Lagage83, Drury83, Droge86}.
Thus, {\it the shock acceleration rate,  $A^\|_{\rm sh}\propto Eu_{\rm
sh}^2/{\bar \kappa}_{ss}\propto E(u_{\rm sh}/v)^2\langle
(1-\mu^2)^2/D_{\mu\mu}\rangle$, is
also a second order mechanism}.
As shown by \cite{Jokipii87}, for oblique shocks
($\theta>0$)  the
acceleration rate also varies as the square of shock velocity, but  in this
case,
specifically for a perpendicular shock ($\theta=\pi/2$) the rate could be
much higher. In general then, as emphasized in \cite{Petrosian12}, in both 
SA by turbulence
and shock acceleration the rates are proportional to the square of the
velocity ratios $u_{\rm sh}/v$ and $v_{\rm A}/v$, respectively, so that the distinction
between them is greatly blurred.
In either process, resonant scattering by turbulence 
provides rapid isotropization of the particle pitch angle distribution,
a necessary prerequisite for efficient acceleration, e.g., \cite{Melrose09}.

More exactly, in the framework of the leaky box model, the
shock
acceleration
rate  can be written as
\begin{equation}
A_{\rm sh}=E\left(1+\frac{1}{\gamma}\right) \left(\frac{u_{\rm
sh}^2}{\bar\kappa_{ss}}\right) \zeta f(\theta, \eta),
\label{Ash}
\end{equation}
where we have introduced the parameter $\zeta=(r-1)/(3r)$ with $r$ the
compression ratio and $f(\theta, \eta)$ which is a somewhat complicated
function of the angle and the ratio of the diffusion coefficients parallel to
perpendicular to the magnetic field $\eta=k_\|/k_\perp$; \cite{Steinacker88,
Jokipii87, Droge86}. For a parallel shock $f=1$ and
${\bar\kappa}_{ss}=\kappa_1+r\kappa_2$, where subscript 1 and 2 refer to
upstream and downstream region of the shock, respectively.
The usual practice is to assume
the Bohm limit; $\kappa\propto vr_g/3$, where $r_g=v/\Omega$ is the gyro radius.
In what follows we will use a more accurate  relation for
${\bar\kappa}_{ss}$ obtained
from wave particle interactions, as those shown in Figure (1). 
For a perpendicular shock the relation again is simple and from \cite{Jokipii87}
we obtain ${\bar\kappa_{ss}}/f(\pi/2, \eta)=2\eta \kappa_{ss}/(1+\eta^2)\sim
2\kappa_{ss}/\eta$ (for $\eta\gg 1$), which amounts to setting 
$f(\theta=\pi/2, \eta)=\eta$. 

\begin{figure}[ht]
\centering
\includegraphics[scale=0.7]{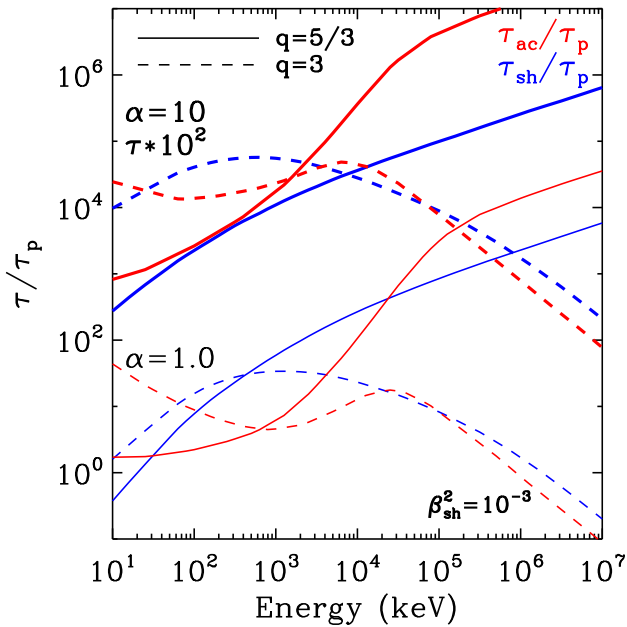}
\caption{Comparison of the SA time (denote here as $\tau_{\rm ac}$) calculated
using Eq. (\ref{times}) and shock
acceleration time (denoted here as $\tau_{\rm sh}\equiv E/A_{\rm sh}$) for
interactions of
electrons with parallel propagating plasma waves
with power law spectral distribution for two indicies $q=5/3$ (Kolmogorov) and
3. Here $\alpha=3.2\times 10^{3}\sqrt{n/{\rm cm}^{-3}}(\mu{\rm G}/B)$, the
ratio of the electron plasma
to gyro frequencies, is a measure of the degree of magnetization, and 
$\tau_p^{-1}\propto \Omega f_{\rm turb}$ is
the characteristic rate for wave-particle interactions. (See more details in
\cite{Pryadko97, Petrosian04}.) 
Note that for highly magnetized plasma ($\alpha=1.0)$, SA is the
dominant mechanism at some energies, and even for a plasma with lower
magnetization, SA cannot be ignored.
The shock velocity is taken to be $u_{\rm
sh}=10^4$ km/s
compared to the Alfv\'en velocity $v_A=c/[(m_p/m_e)^{1/2}\alpha]\sim 7000$ and
700 km/s, resulting in ${\cal M}^2_A=2$ and 200, respectively. The
proportionality
constants
$\zeta$ and $f$ are set to unity. Note that the $q=5/3$ solid curves are
multiplied by
100 for
clarity.
}
\label{SAvsShock}
\end{figure}

{In applications to astrophysical sources we will be dealing with the
scattering and stochastic acceleration times defined as as
\begin{align}
&\tsc(E)=3{\bar\kappa_{ss}\over v^2}=\frac{3}{4}\langle(1-\mu^2)^2/{\bar
D}_{\mu\mu}\rangle,\\
&\tac(E)=\bar
{\kappa}_{pp}^{-1}={p^{2}\over \langle{{\bar D}_{pp}-{\bar D}^2_{p\mu}/{\bar
D}_{\mu\mu}}\rangle},
\label{times}
\end{align}
Using these in Eqs. (\ref{ASA})  and  (\ref{Ash}) we can write shock to SA
acceleration rate ratio as
\begin{align}
\label{rateratio}
\frac{A_{\rm sh}}{A_{\rm SA}}=2\left(\frac{u_{\rm
sh}}{v}\right)^2\left(\frac{\tac}{\tsc}\right) \zeta\xi ''f,
\end{align}
where 
$\xi'' = \gamma^2/(\gamma^2 - 0.5)$. For {\it parallel
shocks} ($f=1$) this ratio becomes $A_{\rm
sh}/A_{\rm SA}\sim (\tac/\tsc))(u_{\rm sh}/v)^2$. As  pointed out above
(and in \cite{Pryadko97}), at low energies
and for
strong magnetic fields, $\tac/\tsc<1$ indicating the dominance of SA. But for
high energies and Alfv\'enic turbulence $\tac/\tsc\sim (v/v_A)^2$
and $A_{\rm sh}/A_{\rm SA}\sim {\cal M}_A^2$, where ${\cal M}_A=u_{\rm
sh}/v_A$
is the Alfv\'en Mach number, so
shock acceleration dominates at high energies and weakly magnetized
plasmas.} Figure \ref{SAvsShock} shows a comparison between  the SA
timescale $\tac$ as defined in Eq. (\ref{times}) and shock acceleration time
$\tau_{\rm ac, sh}=E/A_{\rm sh}\sim(v/u_{\rm sh})^2\tsc$, based on  rates
obtained for
interactions of electrons with parallel
propagating plasma waves \cite{Pryadko97},
for two values of the spectral index  $q$ of the turbulence energy density and 
two degrees of
magnetization described by the plasma parameter $\alpha=3.2\times
10^3\sqrt{n/{\rm cm}^{-3}}(\mu{\rm G}/B)$, which is equal to the ratio of the
electron plasma to
gyro frequencies. As
evident at low energies and small values of $\alpha$ (strong magnetization) SA
is the dominant mechanism. {For oblique shocks the shock rate will be in
general higher by some factor which depends  on the angle $\theta$; e.g. for a
high Mach number ($r=4, \zeta=1/4$) perpendicular  ($\theta=\pi/2$) shock this
factor will be $\sim \eta/2$.} 

In \cite{Petrosian12} it was  suggested that in presence
of a shock the acceleration may be a hybrid process dominated by SA at low
energies and shock at
high energies.
In what follows we will consider the combined processes, which depend on
wave-particle interactions, shock compression ratio and background plasma
parameters.

For a solution of the differential Eq. (\ref{lbox}) we also need the  energy
dependence of the other terms. For the injected spectrum ${\dot
Q}_{\rm inj}(E)$ we will consider a Maxwellian distribution  at a given
temperature $kT\ll m_ec^2$,
and for the energy loss we include ionization and Coulomb losses that dominate
at low energies (and depend on background density $n$ and ionic composition),
and synchrotron and
inverse Compton losses that dominate at high energies (and depend on
the background magnetic
field and
photon energy densities). Coulomb interactions can
also cause energy and pitch
angle diffusion which become important at low energies,
e.g., \cite{McTiernan90, Petrosian08}.

As we will see below the last term, namely the escape time, is the term that can
be obtained most readily from observations, which then allows determination of
the other terms. However, the relation of the escape time to the coefficients of
the
acceleration mechanism is complicated.
As shown in \cite{Petrosian12}, it is related to an integral of spatial
diffusion term
$\kappa_{ss}$ over the acceleration site. Thus, it also depends on the  size $L$
of this site or crossing time $\tcross= L/v$. For the isotropic case with
$\tsc\ll
\tcross$, one expect the diffusion of the particles across the
source to
follow a random walk process, which means we can write   $T_{\rm
esc}\sim
\tau^2_{\rm cross}/\tsc$. In
the opposite limit,  
$\tcross\ll \tsc$, the escape time $T_{\rm esc}\sim \tcross$.  Combining these
two
cases in the past we (\cite{Petrosian04}) have used the  approximate expression
\begin{equation}
\label{tescT}
T_{\rm esc}=\tcross(1+\tcross/\tsc).
\end{equation}
However, other geometric effects such as those produced by the large scale
magnetic fields {e.g. chaotic field lines, or strongly converging or
diverging field configurations (see \cite{ChenQ13})}, or deviation from isotropy
or a simple spherical homogeneous
acceleration site,  can
make the relation between  $T_{\rm esc}$ and other acceleration coefficients
more complex.

\section{The Inversion Process}
\label{inv}

\subsection{The Knowns and Unknowns}

Solution of Eq. (\ref{lbox}) requires knowledge of
energy and {time
dependences} of
the five coefficients involved in the terms on the right side. In situations
where there exist time resolved observations one needs to solve for the time
dependence of the accelerated spectrum. However, if the
dynamic
time $\tau_{\rm dyn}$ describing the evolution is longer than
the characteristic timescales associated with these coefficients (such as
$\tsc, \tcross$ and  $\tac$ or energy diffusion time
$\tau_{\rm diff}\sim E^2/D_{\rm EE}$), then one can use the steady state
assumption and set ${\partial N \over \partial t}=0$ and modulate the results
with the time profile of the dynamic process. This was the case in our
application of the inversion method to solar flares. In the opposite
situation of short dynamic time, and in the absence of temporally resolved
observations, one can integrate Eq. (\ref{lbox}) over the dynamic time in which
case 
$\int_0^\infty {\partial N \over \partial t}dt=0$, because we expect
$N(t=\infty, E)=N(t=0, E)=0$ for high energy  particles. In this
case  one is dealing with the values of the coefficients  averaged over
duration $\Delta t$ of the process; e.g average  injected spectrum  ${\dot
{\bar Q}}(E)=\int_{\Delta t} {\dot Q}(t,E)dt/\Delta t$.%
\footnote{For the sake of simplicity we  shall not use the superscript bar in
what follows.}
As
discussed below this will be the case for the application to SNRs  and cosmic
ray electrons (CRes). Thus, we need to consider only the energy dependence of
the
coefficients.

Two of
these, namely $\dot E_{\rm L}$ and ${\dot Q}_{\rm inj}$, depend on background
plasma parameters
$n, T, B$ and $u_{\rm ph}$, and are independent of the acceleration
process. We will assume that we have sufficient information on the background
plasma so that we know the values and energy dependences of these two terms.
The other three are
related to the characteristics of the acceleration mechanism that we want to
determine. One of these is the energy diffusion coefficient $D_{\rm EE}$
(related to $D_{pp}$). The
escape
time depends on the size $L$ of the source and  on the spatial diffusion rate
(related $D_{\mu\mu}$). The
final term, namely the direct acceleration rate has contribution from
turbulence,
which is related to $D_{\rm EE}$, and from shocks, which is related to
$D_{\mu\mu}$
and the characteristics of the shocks described above.
Assuming that
we know the
value
of the latter and the size $L$ of the acceleration site, we are left with two
primary unknowns $D_{\mu\mu}$ and $D_{pp}$, or in terms of more directly
unknowns $D_{\rm EE}$ and $T_{\rm esc}$. Therefore, in order to
determine the energy dependences of these two coefficients, we need the
variation with energy of two independent observed quantities, as described
next. 

\subsection{Escape Time}

As described in \cite{Petrosian10}, one of the two functions that observations
can provide is
the (spatially integrated) energy spectrum of
the
accelerated particles $N(E)$, which can be deduced from the observed
total photon spectrum $I(\epsilon)$ produced in the acceleration region.%
\footnote{$I(\epsilon)\propto\int dE\sigma(E, \epsilon) vN(E)$, where
$\sigma(E, \epsilon)$ stands for the radiative cross section.}
If the escape time
is finite then the rate of 
particles  escaping
will be 
${\dot Q}_{\rm esc}=N/T_{\rm esc}$. If this spectrum is measured
directly, we then can obtain the escape time simply as 
\begin{equation}
\label{tesc01}
T_{\rm esc}(E)=N(E)/{\dot Q}_{\rm esc}(E).
\end{equation}
If we assume that  Eq. (\ref{tescT}) is an accurate description of how particles
escape, we can then obtain also the scattering time
\begin{equation}
\label{tsc}
\tsc=\frac{\tau_{\rm cross}^2}{N(E)/{\dot Q}_{\rm esc}(E)-\tcross}.
\end{equation}
This will then give a measure of  the pitch angle diffusion coefficients
$D_{\mu\mu}$ and as shown by Eqs.(\ref{Ash}) and (\ref{times}), it will also
give
the acceleration rate by the shock $A_{\rm sh}(E)$, assuming we know
$\zeta$.

\subsection{Energy Diffusion and  Acceleration Rates}

Given the above information we are left with  only two related  unknowns, namely
the energy diffusion coefficient $D_{\rm EE}$ or the direct SA rate $A_{\rm
SA}=2{\bar \xi}D_{\rm EE}/E$. This final unknown can be obtained using our
knowledge of
the
accelerated particle spectra $N(E)$ and the escape time $T_{\rm esc}(E)$ by the
inversion of the
leaky box Eq. (\ref{lbox}) as follows.

The key aspect here is to recognize that this ordinary differential equation
is only first-order in the derivative of $D_{\rm EE}$ with respect to $E$, 
instead of second-order that appears to be the case in its alternate form.
Thus, by utilizing the relation between $A_{\rm SA}$ and $D_{\rm EE}$ in Eq.
(\ref{ASA})
we can rewrite the steady state leaky box equation as
\begin{align}
\frac{d}{d E}
\left[D_{\rm EE}\frac{N}{E}\left( \frac{d \ln{N}}{d \ln{E}}-\xi \right)\right]
+\frac{d}{d E}\left[(\dot{E}_{\rm L}-A_{\rm sh})N\right] \nonumber\\
=\frac{N}{T_{\rm esc}}-{\dot Q}_{\rm inj},
\end{align}
Integrating this  from $E$ to $\infty$ gives 
\begin{eqnarray}
\label{DEE}
D_{\rm EE}=E
\left[ \dot{E}_{\rm L}-A_{\rm sh}+
\frac{1}{N}
\int_E^\infty \left(\frac{N}{T_{\rm esc}} -
{\dot Q}_{\rm inj}\right)dE\right] \nonumber\\
\times\left(2{\bar \xi}- \frac{d \ln{N}}{d \ln{E}} \right)^{-1}, 
\end{eqnarray}
from which we obtain $A_{\rm SA}= 2{\bar \xi}D_{\rm EE}/E$.
Thus, all
the 
terms on the right-hand side can in principle be obtained directly from
observables.
  
Note that for the time integrated equation under consideration here we must have
the
equality $\int_0^\infty  {\dot Q}_{\rm inj}(E)dE=\int_0^\infty N(E)/\tesc(E)dE$.
But
for relevant energies of $E\gg kT$ only a number of particles in
the Maxwellian tail  contribute and   ${\dot Q}_{\rm inj}\ll N(E)/\tesc(E)$. If
this
were not
true  there would be very few particles accelerated and the case
is uninteresting.
Thus, in
what follows we can neglect the injection term. However, given the temperature
of the
background particles this term can be easily calculated and included in the
results.

Finally  we define an
effective acceleration rate as 
\begin{align}
\label{Aeff}
A_{\rm eff}\equiv A_{\rm sh}+A_{\rm SA}\eta_{\rm acc}=  \dot{E}_{\rm
L}+\frac{1}{N}
\int_E^\infty \frac{N}{T_{\rm esc}}dE,
\end{align}
where
$\eta_{\rm acc}\equiv1+ \delta_{\rm acc}/(2{\bar \xi})$,
and we have introduced the spectral index of the accelerated particles
$\delta_{\rm acc}=-d \ln{N}/d
\ln{E}$. At relativistic energies ${\bar \xi}\rightarrow 1$ and, as we will see
below, typically
$\delta_{\rm acc}\sim 2$, so this
rate is sum of the shock and (about two times) SA rates.

\subsection{Escaping Particles}
\label{secEscape}

Escaping particle are measured directly or by the detection of the  radiation 
they produce outside the
acceleration site, which we will call the transport region, where their spectrum
is modified due to
transport effects.%
\footnote{For clarity in what follows the quantities in the transport region is
identified by the superscript ``tr" and those in the acceleration site by
sub- or super-scripts ``acc".}
These effects can be treated by a similar  kinetic equation without the
diffusion and
acceleration terms. If the
particles are injected into a finite region and if one can neglect further
acceleration and assume that pitch angle scattering quickly isotropizes the
particle distribution, then  the evolution of particles in the transport region
can be described by the leaky box Eq. (\ref{lbox}) which now has only the energy
loss and escape terms. Instead of a thermal background source term, the
spectrum of  particles {\it injected  in the transport region} is same as those
escaping the acceleration site;
\begin{equation}
\label{qtr}
{\dot Q}^{\rm tr}_{\rm inj}(E)={\dot Q}_{\rm esc}(E)=\frac{N_{\rm
acc}(E)}{T_{\rm esc}^{\rm acc}(E)}.
\end{equation}
In application to the transport of the \underline{CRs in the
galaxy} we are
dealing with a long dynamic time so that we can use the steady state equation,
which has the
formal solution giving the
effective spectrum of particles integrated over the transport region 
\cite{Stawarz10},
\begin{align}
\label{neff}
N_{\rm eff}(E) = {\tau_{\rm L}^{\rm tr}(E) \over E} \, \int_{E}^{\infty} dE'
\,\, {\dot Q}^{\rm tr}_{\rm inj}(E') \nonumber\\ 
\times\exp\left[- \int_{E}^{E'} {dE'' \over E''}
\,\,
{\tau_{\rm L}^{\rm tr}(E'') \over T_{\rm esc}^{\rm tr}(E'')} \right] \,,
\end{align} 
where  we have defined the energy loss time $\tau^{\rm tr}_{\rm L}\equiv E/{\dot
E}^{\rm tr}_{\rm
L}$. 

Of special importance, in general and in particular for the applications
described below, is the case when the particles escaping the acceleration site
lose all their energy in the transport  region. This is referred to as the {\it
thick target} or {\it totally cooled} spectral model, where one sets
$T_{\rm esc}^{\rm tr}=\infty$
and get a  simpler integral solution
\begin{align}
\label{thicktarget}
N_{\rm eff}(E) &= \frac{\tau_{\rm L}^{\rm tr}}{E} \int_{E}^{\infty} dE'{\dot
Q}^{\rm tr}_{\rm inj}(E')\nonumber\\
&=\frac{\tau_{\rm L}^{\rm tr}}{E} \int_{E}^{\infty} dE'
\frac{N_{\rm acc}(E')}{T_{\rm esc}^{\rm acc}(E')}.
\end{align}
First,  differentiating this equation we derive the desired expression for the
escape time as
\begin{align}
\label{tescThick}
T_{\rm esc}^{\rm acc}(E)={\tau_{\rm L}^{\rm tr}}\left[\frac{N_{\rm acc}}{N_{\rm
eff}}\right]\eta_{\rm eff}^{-1},
\end{align}
where $\eta_{\rm eff}\equiv \delta_{\rm eff}+\frac{d\ln {\tau}^{\rm tr}_{\rm
L}}{d\ln E}-1$,
and we have defined the spectral index $\delta_{\rm
eff}=-d\ln N_{\rm eff}/ d\ln E$.
Second, we note that this last integrand is identical to  the third term inside the square brackets on the
right-hand side
of Eq. (\ref{DEE}), so that with 
the help of this equation we can derive a new simpler relation for the energy
diffusion
rate as 
\begin{eqnarray}
\label{DEEThick}
D_{\rm EE}={E^2}
\left(\frac{1}{\tau_{\rm L}^{\rm acc}} - 
{A_{\rm sh}\over E}+ \frac{N_{\rm eff}}{\tau_{\rm L}^{\rm
tr}N_{\rm acc}}\right)
(2{\bar \xi} + \delta_{\rm acc})^{-1},
\end{eqnarray}
where $\tau^{\rm acc}_{\rm L}\equiv E/{\dot E}_{\rm L}$ is the energy loss time
scale averaged over
the acceleration region.
Finally, we define the effective acceleration time (a combination of shock and
SA
times)
\begin{align}
\label{teff}
\tau_{\rm ac, eff}(E)\equiv \frac{E}{A_{\rm eff}}&=\left[{1\over \tau_{\rm
L}^{\rm
acc}} + {1\over \tau_{\rm L}^{\rm tr}}{N_{\rm eff}\over N_{\rm
acc}}\right]^{-1}\nonumber\\
&=\left[{1\over \tau_{\rm L}^{\rm
acc}} + {1\over \eta_{\rm eff}\tesc} \right]^{-1}.
\end{align}
For pure shock acceleration, the acceleration time $\tau_{\rm ac, sh}=\tau_{\rm
ac, eff}$ and for pure SA, the time 
$\tau_{\rm ac, SA}=\tau_{\rm ac, eff}\eta_{\rm acc}$. Note that while the escape
time depends on only
the ratio of effective to acceleration spectra, the acceleration
times
involve both this
ratio and the energy loss time in the acceleration site. 

In the opposite limit when particles lose very little of their energy in the
transport  region, i.e. when $T_{\rm esc}^{\rm
tr}\ll \tau_{\rm L}^{\rm tr}$, which is called the  {\it thin target} model, 
Eq.
(\ref{neff}) simplifies even further to 
\begin{align}
\label{thintarget}
N_{\rm eff}(E) = T_{\rm esc}^{\rm tr}(E){\dot Q}^{\rm tr}_{\rm inj}(E)
=\frac{T_{\rm esc}^{\rm
tr}(E)}{T_{\rm esc}^{\rm acc}(E)}N_{\rm acc}(E),
\end{align}
from which we get
\begin{eqnarray}
\label{tescThin}
T_{\rm esc}^{\rm acc}(E)=T_{\rm esc}^{\rm tr}(E)[N_{\rm acc}/N_{\rm eff}].
\end{eqnarray}
 For the  diffusion coefficient  in this case we have to replace the
last term inside the first pairs of parenthesis on the right-hand side of  Eq.
(\ref{DEEThick}) by $\int_E^\infty
(N_{\rm eff}/T_{\rm
esc}^{\rm
tr})dE/N_{\rm acc}$. {\it In what follows we will consider only the thick
target
case.}

In summary, the above equations show that  one can determine the pitch
angle and momentum diffusion coefficients in the acceleration region directly
from measurements of the particle spectra in the  acceleration and transport
regions. 

As mentioned at the outset, in \cite{ChenQ13} we have demonstrated the power of
the procedure in
application to solar flares. Here we explore the possibility of using the
radiative signatures of SNRs and observed spectra of CRes in
the interstellar
medium (ISM) to determine the characteristics of
the acceleration mechanism in SNRs.        

\section{Applications to Supernova Remnants}
\label{sec_SNR}

It has been the common belief that SNRs are the source of the observed CRs 
(at least up to the knee at $\sim$10$^{15}$ eV) and
recent high energy gamma-ray observations of SNRs have enforced this belief
considerably. If this is true then we can get information on the two functions
required for our inversion process. The observed radiative spectrum of SNRs from
radio to gamma-rays gives the spectrum of the the accelerated particles,
$N_{\rm acc}(E)$, and the observed spectrum of the CRs provides information on
the
spectrum of accelerated particles escaping the SNRs, $N_{\rm eff}(E)$.
Although in principle this information is available for both electrons and
protons, there are only some preliminary solid observations on the radiative
signature of protons in SNRs. Therefore, in what follows we will focus on the
acceleration of electrons. 

However, it should be emphasized that the
situation here is not as straightforward as in solar flares where these two
functions are determined simultaneously for individual flares. Here we need
knowledge of the transport to the Earth of the electrons escaping the SNRs,
and a more important complexity is that, many and a diverse set of SNRs,
resulting
from explosions of different progenitor stars in different environments,
contribute to the CRs in the ISM. We will address these complexities in the
following sections.

\subsection{Spectrum of Accelerated Electrons in SNRs}
\label{SNRspectra}

Many SNRs are
observed optically and at radio. The radio radiation  produced via the
synchrotron mechanism provides the
original indication of presence of  electrons with energy $E>$ GeV in a magnetic
field of $B_{\rm snr}\sim$ 10--20 $\mu$G.%
\footnote{Note that for extreme relativistic electrons of interest here the
terms
$(1+\gamma^{-1}), {\bar \xi}, \xi', \xi''$ appearing in the above equations are
equal
to one.}
Several SNRs are detected at X-rays which also are
attributed to synchrotron radiation by more energetic electrons, perhaps in a
stronger magnetic field. Fermi and HESS have detected GeV and TeV gamma-rays in
several SNRs.  In some cases, for example SNR RXJ1713.7--3946,  a pure leptonic
scenario, whereby the gamma-rays  are produced by the synchrotron emitting
electrons via the inverse Compton (IC) scattering of cosmic microwave background
(CMB) 
or other soft photons,
seems to work \cite{LiH11}. While in others, e.g., SNR Tycho \cite{Giordano12},
the
hadronic scenario,
whereby the accelerated protons are
responsible for the gamma-rays, fits  the data better. In some others, e.g., SNR
Vela
Jr. \cite{Tanaka11}, both models give acceptable fits.
In any case the
radio and X-ray emission gives information about the spectrum of the {\it
accelerated electrons} which is what we will be concerned with here. We call
this spectrum  $N^{\rm snr}_{\rm acc}(E)$.  

In the case
of solar flares, where nonthermal electron  bremsstrahlung produces the hard
X-ray radiation, one can
use {\it regularized inversion}  procedures to determine the spectrum of the
radiating electrons 
non-parametrically and directly from photon count spectra \cite{Petrosian10}.
Unfortunately this technique cannot be used for
SNRs.
There has not been much effort in
inverting  synchrotron and IC  spectra to obtain electron spectra
non-parametrically. Some time ago, \cite{Brown83} addressed the inversion
of
synchrotron spectra and recently  \cite{LiH11} used a matrix inversion method
of \cite{Johns92} 
to invert the IC spectra and applied it to SNR  RXJ1713.7--3946. But, in
general, 
most of the information on $N_{\rm acc}^{\rm snr}$ is obtained by FF  of the
observed photon
spectra to
parametric electron spectra, with the result that the accelerated electron
spectra  (integrated over
the acceleration region of SNR) {can be described by a power low with a
high energy exponential cut off at energy $E_{\rm snr}$. Here and in what
follows we express all particle energies in units of a fiducial energy $E_0$,
which we set equal to 100 GeV for numerical purposes. Thus, the spectrum of SNR
can be written as} 
\begin{align}
\label{Nsnr}
N_{\rm acc}^{\rm snr}(E)dE=N^{\rm snr}_0f(E/E_0)dE/E_0,
\end{align}
where
\begin{align}
\label{snrspec}
f(x)=x^{-\alpha_1}\exp\left[{-(x/{\rm x_{\rm snr}})^{\alpha_2}+(1/x_{\rm
snr})^{\alpha_2}}\right],
\end{align}
with $x=E/E_0$ and $x_{\rm snr}=E_{\rm snr}/E_0$.
In most cases $\alpha_1\sim 2,\, \alpha_2\sim 0.5$ and  $E_{\rm snr}\sim 6$ TeV
provide
good fits  down to energies of $\sim 2\sqrt{B_{\rm snr}/15\ \mu{\rm G}}$ GeV, 
e.g., \cite{LiH11, Lazendic04}. Note that as defined above $f(1)=1$, and 
$N^{\rm snr}_0$ is
a dimensionless
quantity.

The analyses that lead to the above spectra also indicate presence of
sufficiently strong magnetic field ($B_{\rm snr} \geq 15\ \mu$G) that can come
about from amplification of the weaker  ISM field ($\sim$ 1 $\mu$G) by the
supernova driven forward
shock.
In this case synchrotron losses dominate over IC losses and the radiative loss
time in the acceleration site required for our procedure can be written as 
\begin{align}
\label{tlosssnr}
\tau_{\rm L}^{\rm acc}=\tau_{\rm L,0}^{\rm acc}E_0/E,
\end{align} 
where
\begin{align} 
\tau_{\rm L,0}^{\rm acc} &\equiv \left({6\pi m_ec\over
\sigma_T}\right)\left({m_ec^2\over
E_0}\right)B_{\rm
snr}^{-2}\nonumber\\
&=0.54\times 10^6\left({100\ {\rm GeV}\over E_0}\right)\left({15\ \mu{\rm
G}\over B_{\rm
snr}}\right)^2 {\rm yr},
\end{align}
and $\sigma_T$ is the Thomson cross section.

As mentioned above, however, supernova explosions and SNRs may
have a broad
range of characteristics  and parameters of acceleration. In which
case  the average SNR spectrum contributing to the CRes would depend on
the distribution of the  spectral parameters, 
say $\Phi(\alpha_i, E_{\rm snr})$, where $\alpha_i$ stands for 
$\alpha_1$ and $\alpha_2$. In this case the average spectral shape 
\begin{align}
\label{avsnrspec}
\langle f(x)\rangle=\int\int f(\alpha_i, E_{\rm snr}; x)\Phi(\alpha_i, E_{\rm
snr})d\alpha_idE_{\rm snr}
\end{align}
will depend on the shape of the distribution $\Phi$. As we will
see below
only the value of $\alpha_1$ will be important. This is related to the
power-law indicies of the observed radio spectra which shows a small dispersion
(see \cite {Weiler10}). In addition, as is well known from general
theoretical considerations (\cite{Krymskii77, Axford78, Bell78,
Blandford78}), the power-law index of accelerated particle spectra are
insensitive to
shock characteristics (e.g. compression ratio) for high Mach number shocks, such
as those expected from stellar explosion in the cold ISM. Thus, the  spectral
shape given in Eq. (\ref{snrspec}) seems to be a reasonable
approximation. It should be noted though that explosions and environments of
the upper end main sequence stars are considerably different than those of lower
mass stars (see e.g. \cite{Prantzos86, Woosley02}) and could possibly yield 
different accelerated spectra. Unfortunately there are no observations of
remnants of such stars. This is mainly because they are rarer, which would also
mean they contribute less to CRs. In addition, explosions into a hot stellar
wind environment may lead to a lower Mach number shock and a weaker accelerator.
On the other hand, being more powerful explosions could have an opposite 
effect, which would enhance their contribution.

In the absence of observational evidence about the distribution
of
characteristics of stellar explosions and SNR  spectra, in what
follows we will use the spectral form given in Eq. (\ref{snrspec}) for the
accelerated spectrum $N_{\rm acc}(E)$, with
the cautionary remark that the above unknown may introduce a significant
uncertainty in our final results.

\subsection{Spectrum and Propagation of CR Electrons}
\label{CRspectra}

\begin{figure}[ht]
\centering
\includegraphics[scale=0.42]{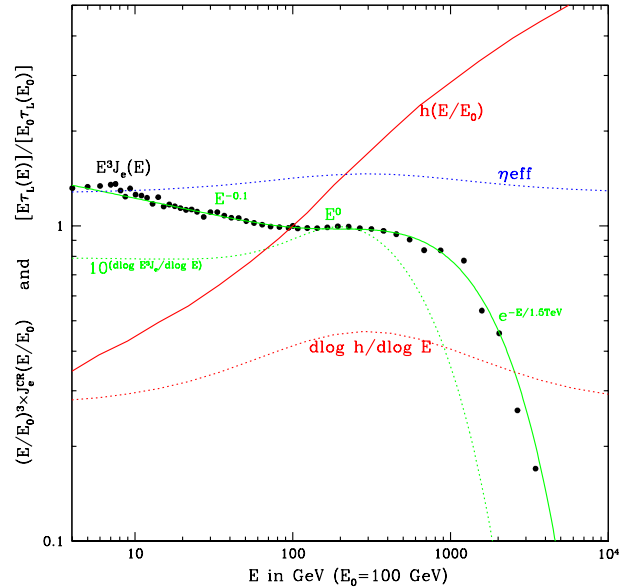}
\caption{Spectrum of CRes multiplied by $E^3$ (dots) as observed by
PAMELA (3 lowest energies), 
HESS (5 highest energies) and Fermi
(from \cite{Adriani11,  Aharonian08, Ackermann10}), respectively. The
solid-green curve gives an
approximate fit to the observations with its logarithmic derivative shown by the
dotted-green curve.   
We also present a sample variation of $h(E/E_0)$ (from \cite{Stawarz10}),
the energy loss time (multiplied by $E$; solid-red) and its
logarithmic derivative (dotted-red) showing the
transition due to the Klein-Nishina effect (see Eq. \ref{tlossgal}).
The dotted-blue curve gives $\eta_{\rm eff}$ defined in Eq. (\ref{tescThick})
and used in Eq. (\ref{tescb}).}
\label{CRespec}
\end{figure}

As mentioned above, it is widely believed that SNRs are  the source of all
CRs, and we will assume this to be
the case for CRes.  Therefore, the spectra of CRs
are related to those of the  particles emitting the SNR radiation via the escape
time from the SNRs. The escaping
particles interact with the galactic background
matter and electromagnetic fields
producing the galactic
diffuse emission from radio to high energy gamma-rays. These interactions and
other processes modify the escaping particle spectra during their transport to
where they radiate and to near the Earth where they are observed directly.
Therefore, CRs are expected to have different spectra than SNRs with the
difference being partially due
to the energy dependence of the escape time and partially due to energy losses
during their transport in the galaxy.
Observations witness these differences. For example, radio spectra of SNRs are
flatter than those of diffuse radio emission in the ISM, and the measured CRe
spectrum $J^{\rm CR}_e(E)$ is different than that given   in Eq. (\ref{Nsnr}).
The spectral flux  of CRes has been measured by many
instruments with varied
results. But most recent measurements by Fermi, HESS and PAMELA have produced a
very precise spectrum shown in Figure \ref{CRespec}. 
As discussed extensively in the literature these spectra
show a well defined deviation from pure power law above 10 GeV and HESS
observations provide a clear evidence of a high energy roll over. 

There has been multiple analyses of this data. Many of these use GALPROP
\cite{Moskalenko98} or
other similar numerical schemes (e.g., Dragon) to account for transport effects
in the galaxy assuming values for background particle and soft
photon densities, large scale magnetic field and a spectrum electromagnetic
field fluctuations.. This is usually carried out by fitting the observed CRe
data to some
parametric form of the spectrum of  {\it the total electrons injected throughout
 the galaxy}, 
which is our function ${\dot Q}_{\rm inj}^{\rm tr}$ (Eq. \ref{qtr}). The results
usually consist
of a primary power law component with index $s$ and a high energy
exponential cutoff%
\footnote{There is also indication of spectral flattening below
4 GeV.
Because of uncertainties due to solar modulation of CRs at such low energies, we
will limit our analysis to energy above 4 GeV.}
at $E_{\rm CRe}$ so that we have
\begin{align}
\label{qinjObs}
{\dot Q}_{\rm inj}^{\rm tr}(E)dE={\dot Q}_{\rm inj, 0}^{\rm tr}g(E/E_0)dE/E_0,
\end{align}
where
\begin{align}
\label{qdotspec}
g(x)=x^{-s}e^{-(x-1)/x_{\rm CRe}}\,\,\,\,\, {\rm with}\,\,\,\,\,  x_{\rm
CRe}=E_{\rm CRe}/E_0.
\end{align}
Here ${\dot Q}_{\rm inj, 0}^{\rm tr}$ is in units of electrons per unit time
and $g(1)=1$.

Different analyses give different explanations for the prominent bump seen 
around
100 GeV. For example, \cite{Ackermann10} attribute
this bump
to a flux of  electrons (plus positrons)  coming from a nearby
pulsar yielding $s=2.7$  and $E_{\rm CRe}\sim 2$ TeV.
\cite{Strong11} 
explain the bump with yet another spectral break, a slight flattening
above 50 GeV and similar values for the other parameters. \cite{DiBernardo13},
using the spectrum of diffuse galactic radio emission, obtain  $s\sim 2.5$  but
do not have the spectral
resolution to see the bump around 100 GeV nor do they see the TeV cutoff. 
We can use the above expression in Eqs. (\ref{qtr}) and (\ref{thicktarget}) to
obtain the
acceleration characteristics. As described below this will be one of the two
methods we will use,
with $s=2.6$ and $E_{\rm CRe}\sim 2$ TeV. 

An alternative and  simpler explanation of the bump in the CRe spectrum
was given in \cite{Stawarz10} (see also \cite{Schlickeiser10}), 
who
show that the energy
dependence of radiative losses due to combined synchrotron and IC scattering (by
star
light, infrared and CMB photons) can account for this deviation. This is because
at
low energies star light is the dominant agent of loss, but at higher
energies IC scattering by star light enters the 
Klein-Nishina (KN for short) regime which suppresses these losses and there is a
transition to
IC losses to infrared and CMB photons  (which are still in the Thomson
regime up to energies of a few TeV) and/or synchrotron losses (depending on
the value of the magnetic field). For typical
values of the relevant quantities in the solar neighborhood
this transition occurs near the bump seen in the CRe spectrum. This
means that in this case the radiative loss time  that enters 
Eq.
(\ref{tescThick}) does not have the simple Thomson regime form
$\tloss\propto E$, but involves an additional function $h(E/E_0)$
that slowly varies with energy in the range from 1 GeV to 1 TeV shown in Figure
\ref{CRespec} (taken
from Fig. 1 of \cite{Stawarz10}).%
\footnote{The initial rise at the lowest energies is due to contribution from
Coulomb collisional
losses.}
The energy loss time in the transport region can then be written as
\begin{align}
\label{tlossgal}
\tau_{\rm L}^{\rm tr}=\tau_{\rm L,0}^{\rm tr}(E_0/E)h(E/E_0),
\end{align}
where
\begin{align}
\tau_{\rm L,0}^{\rm tr}\equiv \left(\frac{6\pi m_ec}{\sigma_T}\right)
\left(\frac{m_ec^2}{E_0}\right)B_{\rm eff}^{-2}.
\end{align}
Here $B_{\rm eff}=\sqrt{8\pi u_{\rm tot}}\sim$ 7 $\mu$G in the solar
neighborhood, where $u_{\rm tot}$ is the energy
density of all soft
photons plus the magnetic field.%
\footnote{The spectrum of injected
electrons  (i.e. ${\dot Q}_{\rm inj}^{\rm tr}$) 
required in this scenario is a power law with spectral index $s=2.42$ with
cutoff at $E_{\rm CRe}=2.75$ TeV.} In this case the observed  CRe flux
spectrum $J_e^{\rm
CR}$ gives 
directly  the effective spectrum  as
\begin{align}
\label{crflux}
N_{\rm eff}(E)dE &=4\pi V_{\rm CRe} J^{\rm CR}_e(E)dE/c \nonumber\\
&\equiv N^{\rm CR}_0 j(E/E_0)dE/E_0,
\end{align}
where $V_{\rm CRe}$ is the volume of the galaxy filled with  CRes,
$j(1)=1$, and $N^{\rm CR}_0=4\pi V_{\rm
CR} J^{\rm CR}_e(E_0)/c$ is the (dimensionless) effective total electron number
at $E_0$. As described below we will use the above
two equation, with the exact observed spectrum for $J_e^{\rm CR}(E)$, as a
second method.
It should be noted that here, unlike in the previous method, 
which assumes presence of nearby pulsar, we assume the solar neighborhood is a
typical location in the galaxy, e.g. does not contain an unusual large scale
fluctuation in density, $B$  field or turbulence (see also the discussion
below).

\subsection{The Two Methods in Practice}
\label{caveats}

We have described  two possible methods for inversion of observations to obtain
acceleration
mechanism
characteristics in SNRs. In what follows we discuss how these methods work in
practice.

The  SNR spectrum $N_{\rm acc}^{\rm snr}(E)$, and either the deduced injected
CRe
spectrum ${\dot Q}_{\rm
inj}^{\rm tr}$ or the observed CRe spectrum $J_e^{\rm CR}(E)$ provide the energy
dependence of the two functions $N(E)$ and $N_{\rm eff}(E)$ that we need for
our analysis
but not their normalization which is required for  determining their ratio.
We have already discussed the uncertainty in the spectrum
$N_{\rm acc}^{\rm
snr}(E)$ above. Here  we describe the uncertainty in the normalizations.
This normalization depends not only on  $N^{\rm snr}_0, {\dot Q}_{\rm inj,
0}^{\rm tr}$ and $N^{\rm
CR}_0$, but also on the rate of SNR formation per unit volume ${\dot
n}_{\rm snr}({\bf r}, t)$. Given this rate we can determine the 
averaged density of accelerated  electrons in the galaxy and the rate of
injection of electrons per unit volume in the ISM as
\begin{align}
\label{NaccTot}
&n_{\rm acc}(E, t)\equiv N_{\rm acc}(E, t)/V_{\rm snr}\nonumber\\
&=\frac{1}{V_{\rm snr}}\int_{V_{\rm snr}}d^3{\bf r} \int_0^t{\dot
n}_{\rm snr}({\bf
r},t_b) N^{\rm
snr}_{\rm acc}(E, {\bf r},t-t_b)dt_b,
\end{align}
and 
\begin{align}
\label{QinjTot}
&{\dot q}^{\rm tr}_{\rm inj}(E, t)\equiv {\dot Q}^{\rm tr}_{\rm inj}(E,
t)/V_{\rm CRe}\nonumber\\
&=\frac{1}{V_{\rm CRe}}\int_{V_{\rm
snr}}d^3{\bf r} \int_0^t{\dot
n}_{\rm snr}({\bf r},t_b) {N^{\rm
snr}_{\rm acc}(E, {\bf r},t-t_b)\over T_{\rm esc}^{\rm acc}(E, {\bf
r},t-t_b)}dt_b,
\end{align}
where $t_b$ is the birth time of  SNRs,  and $V_{\rm snr}$,
the volume of the galaxy enclosing
all SNRs is expected to be or less than $V_{\rm CRe}$. However this
difference does not affect our results.

In general, the integrands vary in time and space, but because 
the active age of a SNR, $\tau_{\rm snr}$, is much shorter than other ages, in
particular
the age of the galaxy, only the SNR formation rate averaged over  the past
$\tau_{\rm snr}$ years enters these equations.%
\footnote{This would be more obvious if one changed the integration variable to
$t'=t-t_b$.}
Moreover, because electrons in several GeV to TeV range  lose their energy
quickly, only the quantities  within
the  finite volume of radius $R\sim\sqrt{3\tau_{\rm L}^{\rm tr}/\tau_{\rm
sc}^{\rm
tr}}\lambda_{\sc}^{\rm tr}\sim
1$ kpc around the solar neighborhood  are relevant (here $\lambda_{\sc}^{\rm
tr}=v\tau_{\rm sc}^{\rm tr}\sim  2$ pc at 100 GeV
is the scattering mean free path of CRes in the ISM).%
\footnote{One can also show that $R/L^{\rm tr}\sim \sqrt{\tau_{\rm L}^{\rm
tr}/T_{\rm esc}^{\rm
tr}}\ll 1$, where $L^{\rm tr}$ is the size of the transport region, in this case
the thickness
of the galactic disk as defined by SNRs or CRs.}
Then the injection rate is determined by the value of the integrand of the above
equations averaged over a small volume and short time $t_0-\tau_{\rm snr}<t<t_0$
or nearly for
$t\simeq t_0$,  the current age of the galaxy.%
\footnote{Note that this also implies that only a small number
of SNRs
contribute to the observed CRs indicating that the contribution of rarer
more massive explosion is less important.}
Thus, we can write 
\begin{align}
\label{Nacc2}
N_{\rm acc}(E, t_0)=N_{{\rm acc},0}f(E/E_0)/E_0,
\end{align}
where
\begin{align}
N_{{\rm acc},0}\equiv N_0^{\rm snr}[V_{\rm snr}{\dot n}_{\rm snr}(t_0)\tau_{\rm
snr}],
\end{align}
and
 \begin{equation}
\label{qinj2}
{\dot Q}_{\rm inj}^{\rm tr}(E, t_0)=N_{\rm acc}(E,t_0)/T^{\rm acc}_{\rm
esc}(E,t_0).
\end{equation}
In what follows we suppress the time $t_0$.
 
These results 
assume  that  $f(E/E_0)$ is  the electron spectrum
integrated or averaged over the active life of
the SNRs. And as stressed above, because the number
of accelerated electrons may vary from SNR to SNR, the  normalization
constants also  stand for averaged quantities. For example, given
the distribution function
${\dot \Psi}(N^{\rm snr})$ the integrand in Eq. (\ref{NaccTot}) is ${\dot
n}_{\rm snr}N^{\rm
snr}=\int_0^\infty N^{\rm snr}{\dot \Psi}(N^{\rm snr})dN^{\rm snr}$.

{\bf Method A:} In this method we use the deduced injected spectrum as given by
Eq. (\ref{qinjObs}). Equating this observed spectrum to that in Eq.
(\ref{qinj2}) we obtain the escape time (from SNRs)  as
\begin{align}
\label{escObs1}
T^{\rm acc}_{\rm esc}(E)=T_{{\rm esc},0}[f(E/E_0)/g(E/E_0)],
\end{align}
with
\begin{align}
T_{{\rm esc},0}= N_{{\rm acc},0}/{\dot Q}_{\rm inj, 0}^{\rm tr},
\end{align}
and the effective spectrum as
\begin{align}
\label{NeffObs1}
N_{\rm eff}&=(\tau_{\rm L}^{\rm tr}/ E){\dot Q}_{\rm inj, 0}^{\rm
tr}\int_E^\infty 
g(E/E_0)dE/E_0 \nonumber\\
&=(\tau_{\rm L}^{\rm tr}/E_0){\dot Q}_{\rm inj, 0}^{\rm tr}{\tilde g}(E/E_0).
\end{align}
Here we have defined ${\tilde g}(x)=\int_x^\infty g(x')dx']/x=
g(x)/\eta_g$, where $\eta_g\sim (x/x_{\rm CRe} + s - 0.5)$.
As shown in Eqs. (\ref{DEEThick}) and (\ref{teff}) the diffusion coefficient and
effective acceleration time depend only on the following combination of terms
\begin{equation}
\label{effoveracc1}
{N_{\rm eff}\over \tau_{\rm L}^{\rm tr}N_{\rm acc}}={1\over T_{{\rm
esc},0}}{g(E/E_0)\over  \eta_g f(E/E_0)},
\end{equation}
and, in particular, the effective acceleration time is obtained as 
\begin{equation}
\label{teffObs}
\tau_{\rm ac, eff}(E)=\tau_{\rm L}^{\rm acc}\left[1+{\tau_{\rm L}^{\rm acc}\over
T_{{\rm esc},0}}{g(E/E_0)\over \eta_g f(E/E_0)}\right]^{-1}.
\end{equation}
We can lump all the unknown and poorly known factors that enter in these
equations
into a single parameter
\begin{align}
\label{Ra}
{\cal R}_a =\frac{T_{{\rm esc},0}}{\tau_{\rm L,0}^{\rm acc}}
=\frac{N_0^{\rm snr}[V_{\rm snr}{\dot n}_{\rm
snr}(t_0)\tau_{\rm
snr}]}{\tau_{\rm L,0}^{\rm acc}{\dot Q}_{\rm inj, 0}^{\rm tr}},
\end{align} 
which then gives
\begin{equation}
\label{timesa}
T^{\rm acc}_{\rm esc}(E)=\tau_{\rm L,0}^{\rm acc}\left[{\cal R}_a{f(E/E_0)\over
g(E/E_0)}\right] \eta_{\rm eff}^{-1},
\end{equation}
and
\begin{equation}
\label{timesa1}
\tau_{\rm ac, eff}=\tau_{\rm L,0}^{\rm acc}\left[{E\over
E_0} + {\cal R}_a^{-1}{g(E/E_0)\over \eta_gf(E/E_0)}\right]^{-1}. 
\end{equation}
Thus, both timescales $T^{\rm acc}_{\rm esc}(E)$ and $\tau_{\rm ac, eff}(E)$ can
be expressed in units
of $\tau_{\rm L,0}^{\rm acc}$ (which depends only on the average magnetic field
in the acceleration
region), and their values and the energy dependence of $\tau_{\rm ac, eff}(E)$
vary with  the value of
the
parameter ${\cal R}_a$. Note that in this method the (more uncertain) energy
loss rate in the ISM does not
enter into these results. Its effect is included in deducing the 
injected spectrum from the observed CRe spectrum. In other words, given the
magnetic field in the SNR
acceleration region around the shock the spectra depend only on ${\cal R}_a$
(or $T_{\rm esc,0}$), 
which involves the properties of the SNRs and the normalization of the deduced
injected electrons.

{\bf Method B:} Alternatively, as mentioned above, we can get the effective
spectrum directly from the observed CRe spectrum as $N_{\rm eff}=N^{\rm
CR}_0j(E/E_0)/E_0$, in which case instead of Eq. (\ref{effoveracc1}) we have
\begin{equation}
\label{effoveracc2}
{N_{\rm eff}\over N_{\rm acc}}={N^{\rm CR}_0\over N_{\rm
acc,0}}{j(E/E_0)\over f(E/E_0)},
\end{equation} 
which when substituted
into Eqs. (\ref{tescThick}) and (\ref{teff}) gives the unknown escape and
effective acceleration times as
\begin{align}
\label{tescb}
\tesc={\tau_{\rm L,0}^{\rm acc}\over \eta_{\rm eff}}\left[{\cal
R}_b{f(E/E_0)h(E/E_0)\over (E/E_0)j(E/E_0)}\right]\eta_{\rm eff},
\end{align}
with
\begin{equation}
\eta_{\rm eff}=-d\ln j/ d\ln E + d\ln h/ d\ln E-2,
\end{equation}
and 
\begin{equation}
\label{teffb}
\tau_{\rm ac, eff}=\tau_{\rm L,0}^{\rm acc}\left[{E\over E_0} + {\cal
R}_b^{-1}{(E/E_0)j(E/E_0)\over
h(E/E_0)f(E/E_0)}\right]^{-1},
\end{equation}
where we have defined
\begin{equation}
\label{Rb}
{\cal R}_b\equiv{\tau_{\rm L,0}^{\rm tr} N_{\rm acc,0}\over \tau_{\rm L,0}^{\rm
acc}N^{\rm CR}_0}=
{V_{\rm snr}[N_0^{\rm snr}{\dot n}_{\rm snr}(t_0)\tau_{\rm snr}]\over N_0^{\rm
CR}}\left({B_{\rm snr}\over
B_{\rm
eff}}\right)^2.
\end{equation}

\begin{figure*}[ht]
\centering
\includegraphics[scale=0.4]{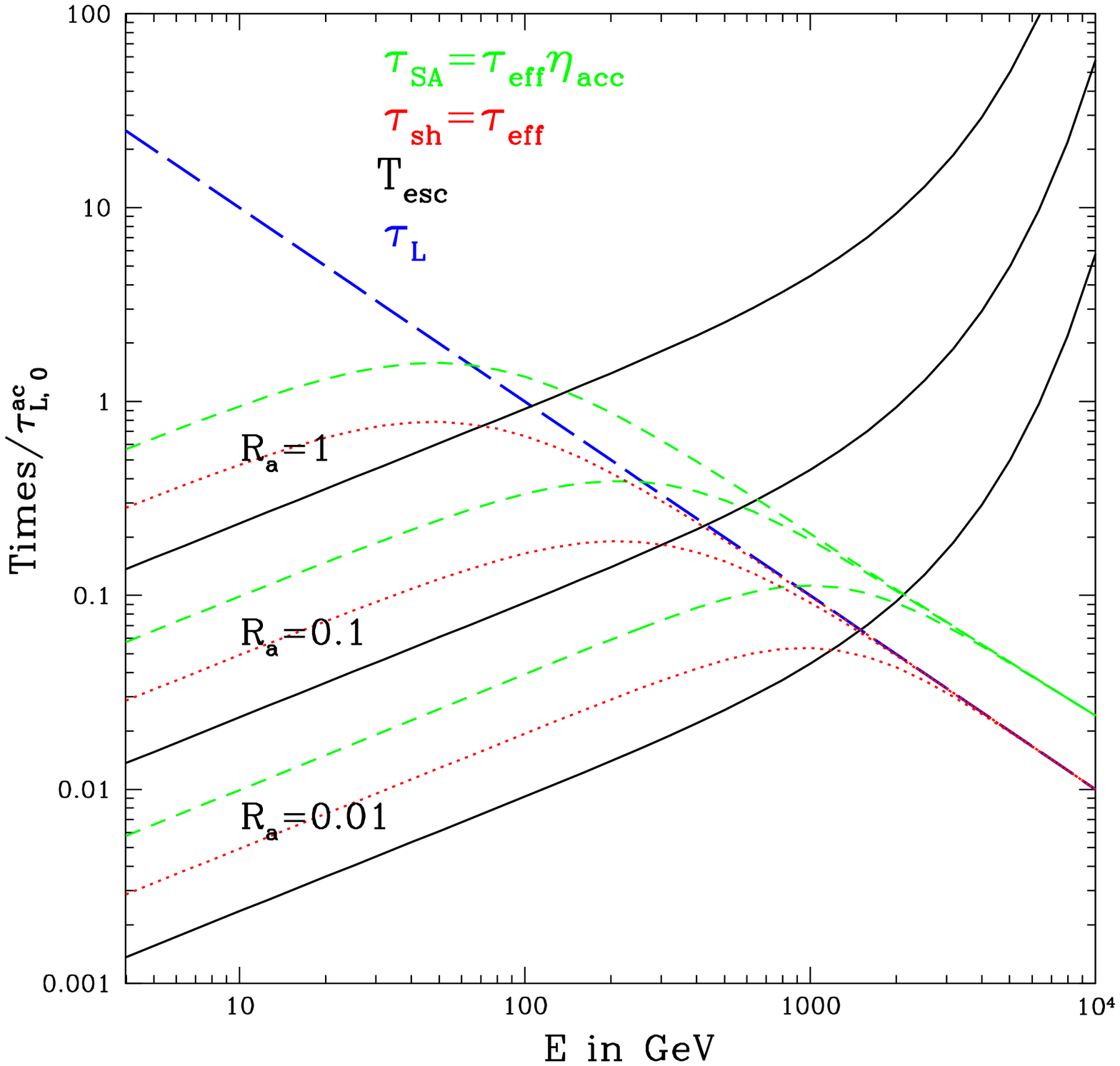}
\includegraphics[scale=0.4]{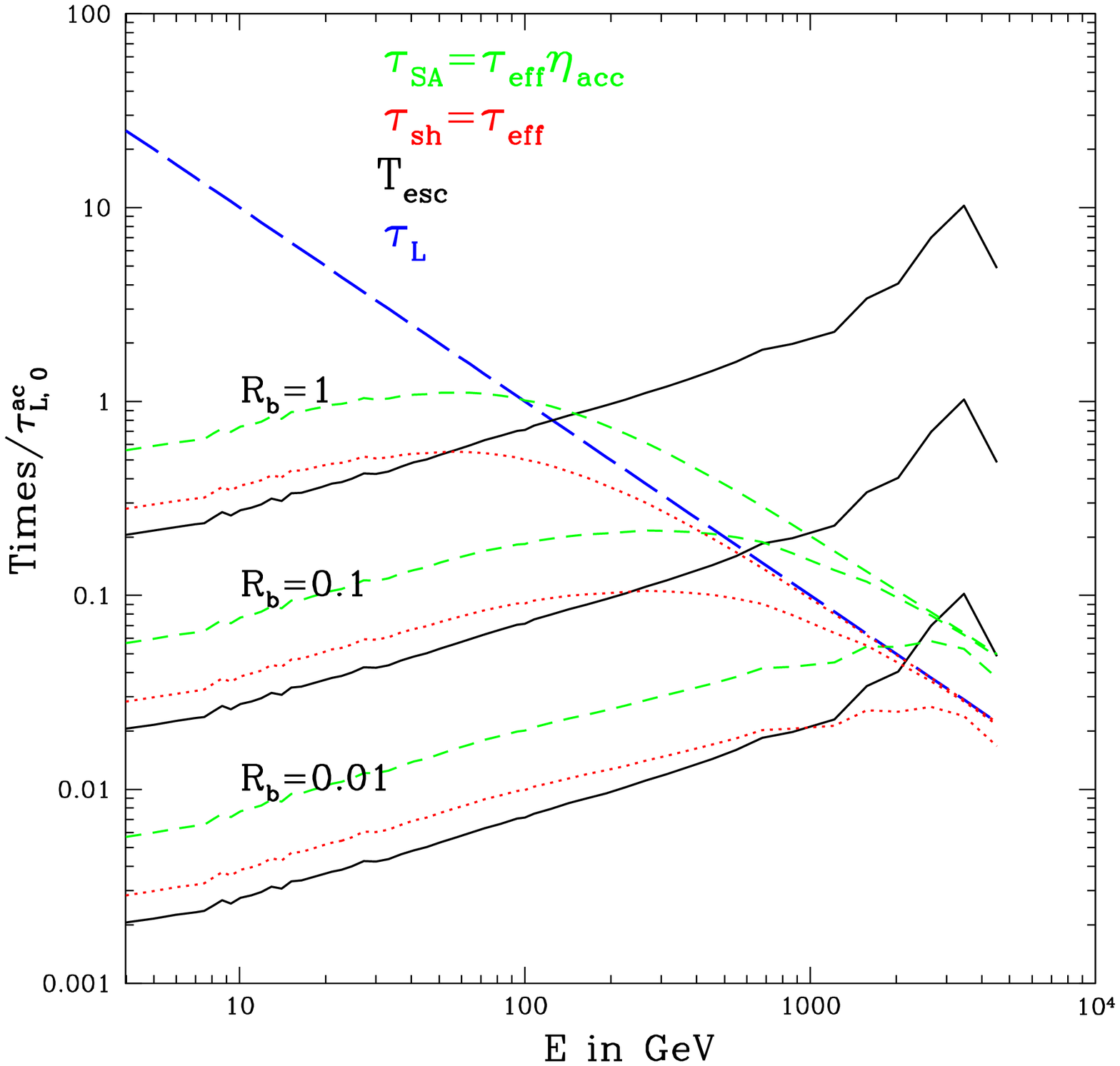}
\caption{Escape, synchrotron loss and acceleration times in SNRs;
black-solid
escape, blue-dashed loss, red-dotted assuming pure shock acceleration
($\tau_{\rm ac, sh}=\tau_{\rm
eff}$) and
green-short dashed  assuming pure SA ($\tau_{\rm ac, SA}=\tau_{\rm
eff}\eta_{\rm acc})$. All times
plotted for three
values
of ${\cal R}_a$, defined in Eq. (\ref{Ra}), are in units of synchrotron energy
loss time at 100
GeV in the acceleration region of the SNR; $\tau_{{\rm L},0}^{\rm acc}\sim
0.5\times 10^6$ yr. {\bf Left:} Based on Method A. {\bf Right:} Based on Method
B.}
\label{fig-timesa}
\end{figure*}

These are very similar to the expressions from Method A but are more directly
related to the
observations
and now the energy loss time in the galaxy comes into play.

Thus, in either method we can combine several poorly understood parameters into
essentially one unknown; namely the constant coefficient ${\cal R}_a$ or ${\cal
R}_b$.
The latter fixes the normalization of the ratio of the effective to accelerated
spectra and determines the relative importance of the two terms that appear in
the expressions for
$\tau_{\rm ac, eff}$ in Eq. (\ref{teff}).

\subsection{Results}
\label{sec_results}

As  mentioned above there is uncertainty
associated with values of the spectral indicies and energy cutoffs. In what
follows we will set $\alpha_1=2, \alpha_2=0.6, s=2, E_{\rm CRe}=2$ TeV and
$E_{\rm
snr}=6$ TeV but will comment on the effects of the uncertainties after
presenting the results. Thus, the remaining unknown  is the dimensionless
factors ${\cal R}_a$ and ${\cal R}_b$.
Before proceeding further we
need to  
estimate their values. Considering the relations between the injection rate
deduced
from the observations and the observed CRe spectrum, it is clear that $N_0^{\rm
CRe}\sim \tau_{\rm
L,0}^{\rm acc}{\dot Q}_{\rm inj, 0}^{\rm tr}$ and that ${\cal R}_a$ and ${\cal
R}_b$ should have similar 
values. Below we estimate their values based on Method B which is more
closely related to the observations. 

There are reliable estimates for the values of the magnetic fields entering in
the expression for
${\cal R}_b$ in Eq. (\ref{Rb}); as stated above $B_{\rm snr}\sim 15\ \mu$G and
using the starlight and
infrared photon densities and magnetic field values in the galaxy one gets
$B_{\rm eff}\sim 7\ \mu$G
e.g., \cite{Stawarz10}. Also using the observed CRe flux (see Fig.
\ref{CRespec}) of $E^3J^{\rm CRe}_e(E)|_{E = 100\ {\rm
GeV}}=120$ GeV$^2$/(s sr m$^2$), we get 
$N_0^{\rm CRe}=5\times\ 10^{-18}\ {\rm cm}^{-3}\ V_{\rm CRe}\sim
3\times 10^{50}$,  assuming the poorly known volume of the galaxy that is filled
with CRes
to be
$V_{\rm CRe}\sim  6\times 10^{67}$ cm$^3$. Even less well known are the values
of the terms in the
square brackets in the numerator of Eq. (\ref{Rb}). The rate of occurrence of
supernovae is believed
to be about several per century but what fraction of these produce active (i.e.
CR producing)
remnants is
not well known. Observations seem to indicate a smaller rate ${\dot n}_{\rm
snr}$. The active age of
SNRs is estimated to be around $10^4$ to $10^5$ yr, which gives a rough estimate
of  $V_{\rm snr}{\dot n}_{\rm
snr}\tau_{\rm snr}\sim 100$. The final factor namely $N_0^{\rm snr}$ can be
estimated from the observed
synchrotron and X-ray radiation intensities of  individual SNRs. For example,
SNR 
RXJ1713.7--3946
has an observed  peak  flux (at X-rays) of $\nu F_{\nu}\sim 600$ eV/(s
cm$^{-2}$) and a low energy
spectrum $F_{\nu}\propto \nu^{-0.5}$.  Assuming a distance of 6 kpc, we get
a good estimate
for
the  total energy of the synchrotron radiation  ${\dot {\cal E}}_{\rm
syn}\sim 4\times 10^{36}$ ergs/s.
This is related to the accelerated particle spectra as 
\begin{equation}
\label{synch}
{\dot {\cal E}}_{\rm syn}=\int_0^\infty N^{\rm snr}_{\rm
acc}(E){\dot E}_{\rm syn}(E)dE,  
\end{equation}
where  ${\dot  E}_{\rm syn} = E^2/(\tau^{\rm acc}_{\rm L,0}E_0)$ is the
synchrotron energy loss rate. For the assumed spectral parameters this gives
${\dot {\cal E}}_{\rm syn}=N_0^{\rm snr}E_{\rm snr}/\tau^{\rm acc}_{\rm
L,0}$
or $N_0^{\rm snr}=6\times 10^{48}$. 

Putting all these together we get ${\cal R}_b\sim 1$. However this is most
likely an overestimation
because we have used the observations from a bright SNR. The number of
accelerated electrons
for an average SNR (including possibly a substantial population of weak and
undetected ones) would
lower this value considerably. For example, using the general belief that
supernovae inject
$10^{51}$ ergs into the ISM and that say 10 percent of this going to CRs, with
an electron share of
one to two percent, we get a number of accelerated electrons smaller by a factor
of 10, or 
$N_0^{\rm snr}\sim 10^{48}$ or 
${\cal R}_b\sim 0.1$.  Considering the large uncertainties about all the
above numbers, 
in what follows we present results for three values of ${\cal R}_a={\cal
R}_b=1.0, 0.1$ and 0.01 spanning a wide enough range to account for all
uncertainties.

Figure \ref{fig-timesa} shows variation with energy of
all time scales obtained by Method A (left) and Method B (right)
normalized to the value of synchrotron energy loss time at 100 GeV in the SNR
($\tau_{\rm L,0}^{\rm
ac}\sim 0.5\times 10^6$ yr). As evident the two methods give very similar
results but Method B results end where the
observations of CRe spectra become unreliable.

\section{Interpretation and Discussions}
\label{sec_interp}

Let us first consider {\bf the escape time} which is essentially the ratio of
the accelerated spectrum to
observed CRe spectrum multiplied by the loss time. At energies below $E_{\rm
CR}=2$ TeV it is nearly a power law with index $\sim (s-\alpha_1)= 0.7$ in
Method A and is $\sim (d\ln j/d\ln E+d\ln h/d\ln E)-(1+\alpha_1)\sim 0.3$
in Method B, with difference primarily due to  the KN effect. $\tesc$  starts
to increase  steeply at $E>$ 1--2 TeV. This rise makes the escape
of high energy electrons from the SNRs more difficult, and is the  causes of the
steep (exponential) decline in the observed CRe spectrum.

{\bf The  acceleration times} for pure shock or pure SA have similar energy
dependences (with a factor
of $\eta_{\rm acc}\sim 2$ difference between them; with SA requiring a
longer time or a lower rate).
At low energies these times are dominated by
the second term in Eqs.  (\ref{teffObs}) and (\ref{teffb}), which
makes them  proportional to the escape time.  Had this trend continued to
higher energies the acceleration time
would have exceeded the energy loss time which would have caused a spectral
cutoff  when these
times would have been equal (e.g., at 0.1, 0.5, and 3 TeV for ${\cal R}_b$ =
1.0, 0.1, and 0.01,
respectively; and at smaller values by a factor of about 2 for Method A). Since
the deduced SNR electron spectra are
observed to cut off at higher energies (6 TeV for RXJ 1713.7), the acceleration
time must decrease to
remain below the energy loss time as seen in both  figures.%
\footnote{Note that the definition of the SA time  is not unique. As
defined here, the SA
times can be  longer than the loss time and still give a power law spectrum
because of the influence
of the energy diffusion term.}

\begin{figure*}[ht]
\centering
\includegraphics[scale=0.4]{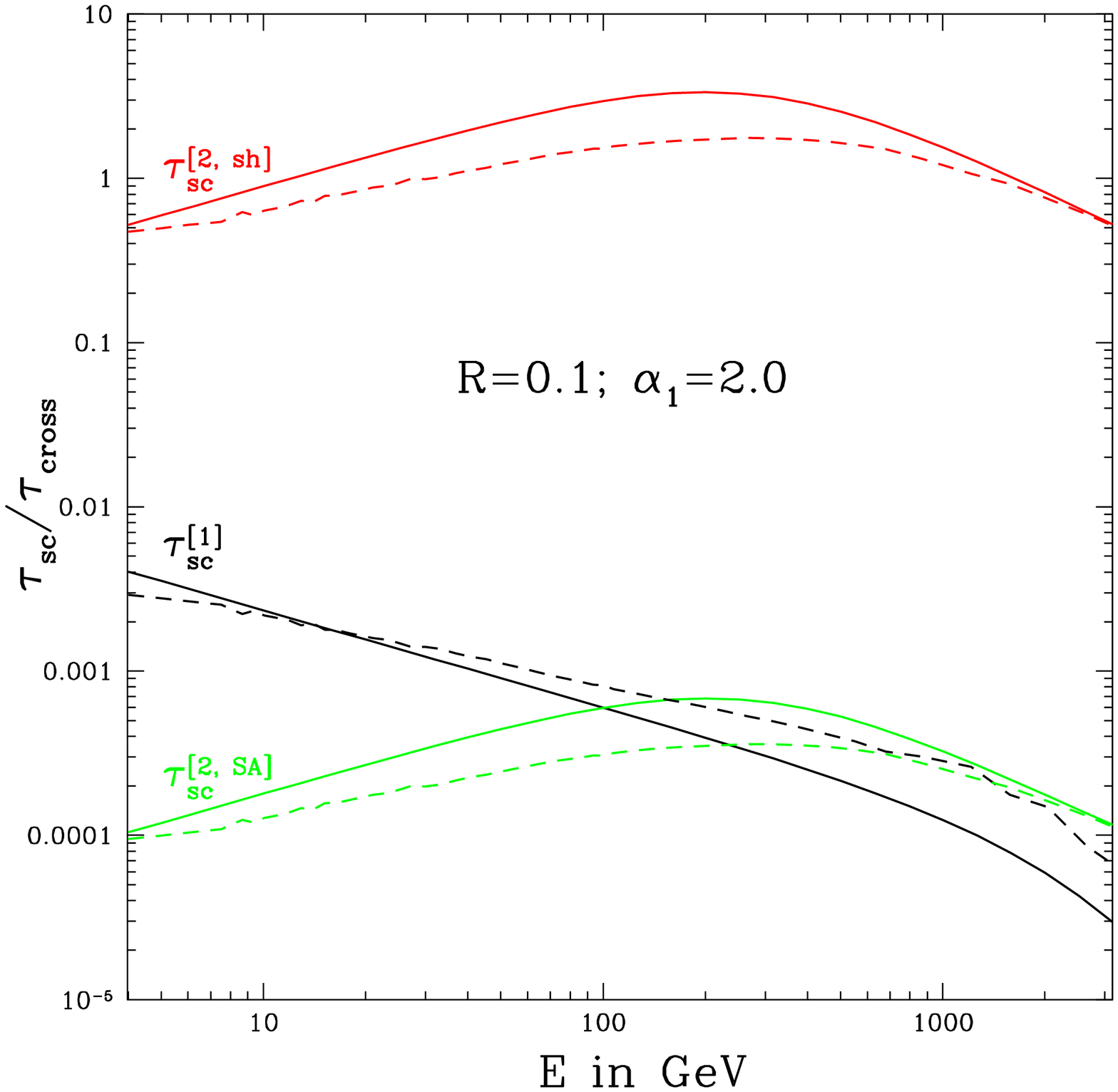}
\includegraphics[scale=0.4]{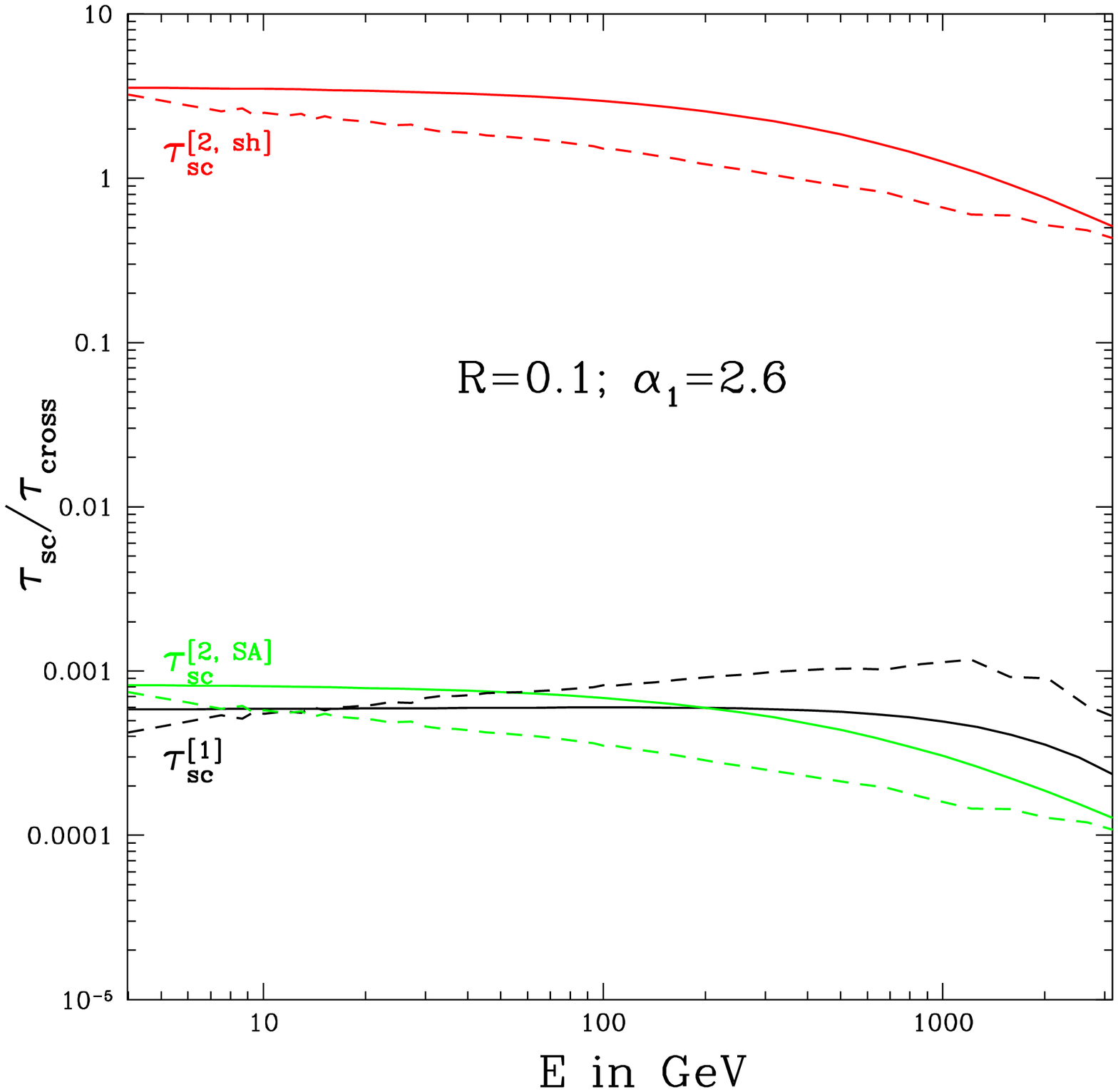}
\caption{Scattering times obtained from the relation between escape and
scattering times in Eq. \ref{tescT} (black), and the relation between the
acceleration and scattering times; Eq. \ref{Ash} for pure shock acceleration
(red) and pure SA from the simple relationship $\tsc^{[2,{\rm SA}]}=\tau_{\rm
acc, SA}\beta_A^{-2}$ (green), valid for relativistic energies and Alfv\'enic
turbulence ($c\beta_A$ is the Alfv\'en velocity). Solid and dashed curves
obtained using Methods A and B, respectively using ${\cal R}=0.1$ for both. {\bf
Left:} For spectral index $\alpha_1=2.0$. {\bf Right:} For spectral index
$\alpha_1=2.6$.}
\label{scattimes}
\end{figure*}

As evident from the discussion in \S \ref{model} and
\ref{inv} we can also
obtain {\bf the scattering time}  in the acceleration site. 
For this purpose we need some information about the background plasma in
the acceleration site. The first is the size $L$ of the region. We will use
the fiducial value of 10 pc (to include the effects of the chaotic
structure of the large scale magnetic field; see Footnote 1) which gives us a
crossing time $\tcross\sim 30$ yr. We also need the shock, Alfv\'en and sound
velocities. We shall assume a shock
velocity of $10^4$ km/s or $\beta_{\rm sh}^2\sim 10^{-3}$, and   Alfv\'en 
velocity of
100 km/s (for $B\sim 15$ $\mu$G, $n=0.1$ cm$^{-3}$) or
$\beta_A^2\sim 10^{-7}$, so that  
the Alfv\'en Mach numbers is very large as one needs for
efficient shock acceleration. For such such high Mach
numbers the
compression ratio $r=4$ and $\zeta = 1/4$.

There are, however, two different ways of obtaining the
scattering time.
The first way, which is common for both shock or SA, comes from the relation
between the  escape and scattering times, which if we assume a  random walk
process of escape  is described by Eq. (\ref{tescT}) and involves the
crossing time. Given that $\tesc> 10^3\ {\rm yr}
\gg \tcross \sim$ 30 yr we obtain the first estimate for the
scattering time as:
\begin{align}
\label{tscObs1}
\tsc ^{[1]}= \frac{\tau^2_{\rm cross}}{\tesc}
\sim 0.025\ {\rm yr}~\left(\frac{L}{10\ {\rm
pc}}\right)^2\left(\frac{4\times 10^4\ {\rm
yr}}{\tesc}\right),
\end{align} 
which as expected is much shorter than the crossing time. Here and in what
follows the numerical values are calculated for $E=100$ GeV and ${\cal
R}=0.1$.

The second method of determining $\tsc$ comes from the
relation between the
acceleration and scattering times. For pure {\bf shock acceleration} $\tau_{\rm
ac, sh}=\tau_{\rm ac, eff}$ and as seen in Eqs. (\ref{Ash}) and
(\ref{times})  the energy dependence of the scattering  and acceleration times  
should be similar but their relative value depends on the shock velocity, the
factor $\zeta$, and for perpendicular shocks on $\eta=\kappa_\|/\kappa_\perp$.
Neglecting the latter for now we get
\begin{align}
\label{tscsh}
\tsc^{[2, {\rm sh}]}=\zeta\left(\frac{u_{\rm sh}}{c}\right)^2\tau_{\rm ac, sh}
\sim 10\ {\rm yr}\left(\frac{\beta^2_{\rm
sh}}{10^{-3}}\right)\left(\frac{\tau_{\rm ac, sh}}{4\times 10^4\ {\rm
yr}}\right),
\end{align}
which is about the crossing time and much larger than the the first estimate  of
scattering time. It also has a different energy dependence. As can be seen in
Figure \ref{scattimes} (left), for the spectral indexes ($\alpha_1=2, s=2.6$)
assumed above   the first
estimate (black curves) decreases monotonically with energy while the second
(red curves) first increases with energy and then declines at higher energies.
The difference in energy dependence at low energies comes from the fact that
here $\tesc\propto \tau_{\rm ac, eff}$ [see Eqs. (\ref{timesa}) and
(\ref{timesa1})] making 
$\tsc^{[2, {\rm sh}]}\propto 1/\tsc^{[1, {\rm sh}]} (\propto E^{\alpha_1-s}$,
for Method A). This difference will be less severe for a steeper SNR electron
spectra (i.e. for $\alpha_1$ closer to $s$), which is the case in some SNRs. For
example, in SNR S1993J with radio spectral index of $\sim 0.8$ one gets 
$\alpha_1\sim 2.6$ (\cite{Weiler10}). As shown in Figure \ref{scattimes}
(right)
using $\alpha_1=2.6$ we get  similar energy dependence for both estimates (and
both methods).

However, as shown above the absolute values of the scattering time deduced
from the two curves are  different by a large factor:
\begin{equation}
\label{ratioSCAT}
\frac{\tsc^{[2, {\rm sh}]}}{\tsc ^{[1]}}=4000\left(\frac{\beta^2_{\rm
sh}}{10^{-3}}\right)\left(\frac{10\ {\rm pc}}{L}\right)^2\left({\tau_{\rm
sh}\tesc\over 1.6\times 10^9\ {\rm yr}^2}\right).
\end{equation}
 Agreement can be obtained for a lower shock
velocity ($\sim  150$ km/s) and/or a larger crossing time  ($L\sim  60$ pc). 
There is more uncertainty in the first of the above
two ways of computing the scattering time; for example, as
mentioned above and in Footnote 2, in a chaotic magnetic field of scale
$\lambda_B
\ll L$ the effective crossing time will be larger  by $L/\lambda_B$, which will
reduce the
above discrepancy by the square of this  factor. Thus, for concordance we
require 
$L^2/\lambda_B\sim 600$ pc; (e.g., $\lambda_B\sim 0.01$ pc for $L=2.5$
pc). 
{As mentioned in connection with Eq. (\ref{rateratio}), for a perpendicular
shock this ratio decreases by the factor
$\eta/2$ expected to be much larger than one so that the required conditions
may not be as extreme.}

More generally,   the validity
of the use of the random walk relation between
escape and scattering times may also be questionable, so that these results may
be
telling us that the relation of the escape time to the scattering and crossing
times is more complicated than given by the random walk hypothesis. For example,
in a near perpendicular shock, where particles spiral up and down the surface of
the shock and  escape when  they are scattered perpendicular to the shock
front, the escape time may be  proportional to the diffusion coefficient
perpendicular to the magnetic field giving 
$\tsc ^{[1]}\propto \tesc/\eta$, which could bring the shape and value of the
first estimate closer to that of the second estimate. To our knowledge there has
not been much discussion of this aspect of the problem in the literature so that
these possibilities require further explorations, which are beyond
the scope of this paper.

{\bf Stochastic acceleration} by turbulence may be important or even
dominant if there is weak or no
 turbulence in the upstream region, conjectured to be  generated by the
accelerated particles. In
this case most of the
acceleration may happen in the downstream turbulent  region with particle
escaping into the ISM once
they cross the shock into the upstream region. However, this mechanism also
faces similar
difficulties. Here the energy dependence of the acceleration
time 
$\tau_{\rm ac, SA} \sim 2\tau_{\rm ac, eff}$ (or
energy diffusion time)
is related to the scattering time via the relation between $D_{pp}/p^2$ and
$D_{\mu\mu}$. In most
wave-particle interaction scenarios these two coefficients have fairly similar
energy dependences especially at relativistic energies. Electrons with energies
above few GeV interact mainly with Alfv\'en or fast mode waves in which case
$D_{pp}/p^2=D_{\mu\mu}\beta_A^2\propto E^{q-2}$ so that 
$\tau_{\rm ac, SA}\propto p^2/D_{pp}$ and $\tau_{\rm sc,SA}\propto
1/D_{\mu\mu}$ 
(see e.g. \cite{Pryadko97}). Thus, we have a second
estimate for scattering time for SA as well:
\begin{align}
\label{tscSA}
\tsc^{[2, {\rm SA}]}=\tau_{\rm ac, SA}\beta_A^2=10^{-2}\ {\rm
yr}\left(\frac{\beta^2_A}{10^{-7}}\right)\left(\frac{\tau_{\rm acc, 
eff}}{10^5\ {\rm yr}}\right).
\end{align}
As shown by the green lines in Figure \ref{scattimes},
in this case also the energy dependences of $\tsc^{[1]}$ and $\tsc^{[2,
{\rm SA}]}$ disagree at low energies for $\alpha_1=2$ (left) but they roughly
agree at high energies, and, again, the agreement is improved for
$\alpha_1=2.6$ (right), where both times have almost a flat energy dependence
requiring a turbulence spectral index of $q=2$, which is somewhat  greater than
the Kolmogorov index. Moreover, now the relative absolute values are in better agreement
for the assumed values of Alfv\'en  velocity of 100 km/s and effective size of
$L\sim 10$ pc.

\section{Summary}
\label{sec_sum}

We consider acceleration of particles in the framework of the leaky box version
of the
Fokker-Planck kinetic equation, which provides an adequate description of the
pitch angle averaged and
spatially integrated (over the acceleration region) energy spectrum of the
accelerated particles.
This equation describes  SA by turbulence and/or acceleration by a shock, where
the leaky box
encloses the upstream and downstream turbulent regions of the shock. Turbulence
plays a central role
in both mechanisms, with the momentum diffusion coefficient $D_{pp}$ determining
the rate of
energy diffusion and acceleration in the SA model, and with the  pitch
angle diffusion  coefficient $D_{\mu\mu}$ determining the spatial
diffusion coefficient $\kappa\sim v^2/D_{\mu\mu}$, hence the rate of
acceleration by the shock. In
addition, the energy loss rate, shock compression ratio (or Mach number) and 
relative values of the spatial diffusion coefficients  parallel and
perpendicular to the magnetic field, and in the upstream and downstream regions,
also come into play. In
the leaky box scenario the coefficients $D_{pp}$ and $D_{\mu\mu}$ are
represented by the energy diffusion coefficient
$D_{\rm EE}$ and the escape time $T_{\rm esc}$ of the particles from the 
acceleration site. Thus, if we can measure the latter two coefficients we can
determine the fundamental wave-particle interaction rates and shed light
on the nature of  turbulence

$\bullet$ As demonstrated in \cite{Petrosian10}, we can obtain the escape time
from the measured spectrum of the accelerated particle $N(E)$ and that of
the escaping
particles $N(E)/\tesc(E)$. We further demonstrate (see \cite{ChenQ13}) that with
 the inversion of the differential kinetic equation into its integral
form,
we can obtain the energy diffusion coefficient {\it
non-parametrically and directly}
from observations of the two spectra and the energy
loss rate of the particles in the acceleration region.

$\bullet$ We also show that the relations between the two unknowns and
observables simplifies
considerably if the escaping particles lose all their energy in the transport
region
outside the acceleration site.

$\bullet$ We demonstrate how this procedure can give us the two unknown
characteristics of the
acceleration mechanism in SNRs using the spectrum of the accelerated electrons
deduced from radio,
X-ray and gamma-ray observations of the SNRs and the observed galactic CRe
spectrum.

$\bullet$ Expressing all the coefficients or rates in terms of their associated
timescales (e.g.,
acceleration and scattering times), we show that the unknown time scales can be
expressed in units of
the relatively well known synchrotron energy loss time in the SNR and a single
parameter which is a
combination of various observable scaling factors, such as rate of formation and
length of active
period of SNRs and other secondary factors.

$\bullet$ We employ two different methods of treatment of the observations and
show the deduced
energy dependence of escape and acceleration times  for some
reasonable value of the parameters, which in principle can
be known given sufficient detailed observations. In Method A we use the spectrum
of injected
electrons  into the ISM deduced  from the observed CRe
spectrum (e.g.,
using  GALPROP or other similar models for transport of electrons in the
ISM). In Method B we use
the observed CRe spectrum directly using a simplified transport
dominated and
IC losses by starlight which is affected by the KN effects as described in
\cite{Stawarz10}.

$\bullet$ For interpretation of the results, we show that we
can
obtain scattering time ($\tsc\sim1/D_{\mu\mu}$) of particles in the
acceleration region using two different relations between  it and the above
timescales. The first is from its relation to the escape time,
which is mediated by the crossing time ($\tcross=L/c$) as $\tsc=\tau _{\rm
cross}^2/\tesc$ assuming a random walk situation when $\tesc\gg \tsc$. The
second is from its relation to the acceleration times. For shock acceleration
scattering and acceleration times are proportional to each other with
proportionality constant being $(u_{\rm sh}/v)^2$ (plus factors $\zeta$ and
$\eta$). For pure SA of greater than few GeV electrons by Alfv\'en or fast mode
waves there is a similar relation but with proportionality constant of
$(v_A/v)^2$.

$\bullet$We find that, for the values of the parameters used in
our
calculation (specifically the spectral index  $\alpha_1=2.0$), the
two estimates of the scattering time  give  very different energy dependences
for the scattering time. This discrepancy largely  disappears for
$\alpha_1=2.6$. Given the caveats stressed in our discussion this is not an
unlikely resolution of the problem.

$\bullet$ Assuming
presence of a sufficient intensity of turbulence both in the upstream and
downstream
regions of the shock, we expect the acceleration in SNRs to be dominated by the
shock, because of the
prevailing high Mach numbers.  However, for this scenario we find that the 
absolute values obtained
by the two relations are  different by a factor of about 1000 for our
fiducial values of $10^4$ km/s for shock velocity and $L=10$ pc for size. This
discrepancy will be smaller for a perpendicular shock. This lead us to our {\it
first conclusion that, in addition to a steeper spectrum for accelerated
electrons, either these values are off by an order of magnitude,  or that the
escape time is not related to the crossing time in the
simple way  one obtains from the random walk scenario. The latter is an
important result and needs further exploration.} 

$\bullet$ On the other hand, in absence of a sufficient
intensity of turbulence in the upstream region, whose presence is only
conjectured and not  established definitely yet, one can have a pure
SA of  particles in the turbulent downstream region. It turns out that in this
scenario  the absolute values of the two scattering times roughly agree. {\it
This, lead us to the  second conclusion that in the  SA scenario having a
steeper accelerated electron spectrum is sufficient and it requires a spectrum
of turbulence that is slightly steeper than Kolmogorov.}

These are clearly preliminary results, but they demonstrate the power of the
inversion method
developed here. A
more detailed analysis of the existing data on emission from SNRs and transport
of the CRes
can provide better values and forms for the observables required for the
inversion, and a more
detailed analysis of the inversion, e.g., including time dependence, can 
constrain the models
further. These will be addressed in future publications. But we can conclude
that the above results
indicate that either the spectrum of injected electrons in the ISM deduced from
CRe and
galactic diffuse emissions (Eq. \ref{qinjObs}) is incorrect and/or the simple
relation between
escape and scattering times used assuming the random walk scenario is incorrect.
The
latter is more likely
to be the case and is similar to
the conclusion we reached  applying these techniques to solar flares. There
mirroring of
electrons in a converging magnetic field configuration was invoked to resolve a
similar discrepancy.
Perhaps a complex large scale field geometry can help in SNRs as well. On the
other hand, more consistent results are obtained for a pure
stochastic acceleration scenario.

\end{document}